\documentclass[nofootinbib,aps,11pt]{revtex4}
\pdfoutput=1
\usepackage{graphicx}
\usepackage{cancel}
\usepackage{amssymb}
\usepackage{textcomp}
\usepackage{amsmath}
\usepackage{bm}
\usepackage{times}
\usepackage{epsfig}
\usepackage{color}
\usepackage{graphics}
\usepackage{hyperref}
\usepackage{setspace}

\def\nn{\nonumber}

\def\mm{\mathcal{M}}
\def\BR{\text{BR}}

\def\squark{\tilde{q}}
\def\squarka{{\tilde{q}_{\alpha}}}
\def\qqbar{q\bar{q}}

\newcommand{\mxone}{m_{X_1}}

\newcommand{\ZBL}{Z_{BL}}
\newcommand{\mzbl}{m_{Z_{BL}}}
\newcommand{\ABL}{A_{BL}}
\newcommand{\mabl}{m_{A_{BL}}}

\newcommand\plot[2]{\includegraphics[width=#1, angle=0]{plots/#2}}
\newcommand\Fdiag[2]{\includegraphics[#1]{diags/#2}}

\newcount\hour \newcount\minute
\hour=\time \divide \hour by 60
\minute=\time
\begin{document}

\title{\Large {The LSP Stability and New Higgs Signals at the LHC}}
\author{Pavel Fileviez P\'erez}
\email{fileviez@physics.wisc.edu} 
\author{Sogee Spinner}
\email{sspinner@wisc.edu}
\author{Maike K. Trenkel}
\email{trenkel@hep.wisc.edu}
\affiliation{Phenomenology Institute, Department of Physics, 
\\
University of Wisconsin-Madison,
1150 University Avenue, Madison, Wisconsin 53706, USA}
\date{\today}

\begin{abstract}
\vspace{1.0cm}
The fate of R-parity in the context of the minimal supersymmetric standard model is a central issue 
which has profound implications for particle physics and cosmology. In this article we discuss 
the possibility of testing the mechanism responsible for the stability of the lightest supersymmetric 
particle at the Large Hadron Collider (LHC). The simplest theoretical framework where R-parity 
conservation can be explained dynamically allows for two types of B-L models. In the first scenario the 
new Higgses decay mainly into two right-handed neutrinos giving rise to exotic lepton number violating 
signals together with displaced vertices. In the second model one could have peculiar channels 
with multileptons and/or multiphotons in the final state. In both cases, the local B-L gauge 
symmetry is broken at the TeV scale and the discovery of the new Higgs bosons may be possible at the 
LHC. We investigate in detail the production mechanisms for the Higgs bosons relevant for the LHC 
and the key decays which would shed light on how R-parity is conserved. These results may help to 
understand the link between the cold dark matter of the universe and the missing energy that could be 
observed at the LHC if supersymmetry is realized in nature.          
\end{abstract}

\maketitle

\newpage

\begin{spacing}{1.5}
\tableofcontents
\end{spacing}

\newpage


\section{Introduction}

The main goal of the Large Hadron Colllider (LHC) is to discover the mechanism responsible for electroweak symmetry breaking 
in the context of the standard model (SM) or in a new TeV scale theory. The minimal supersymmetric standard model (MSSM) 
is considered as one of the most appealing contenders for this new theory. In this context two important cosmological 
issues can be solved: the matter-antimatter asymmetry can be understood through the electroweak baryogenesis mechanism 
and the cold dark matter of the universe candidate may be the lightest supersymmetric particle (LSP). See Ref.~\cite{SUSY} 
for a review on phenomenological and cosmological aspects of supersymmetry. 

The fate of R-parity in the context of the MSSM is a central issue which has profound implications for particle physics 
and cosmology. R-parity is defined as $R=(-1)^{2S} M$, where S and $M=(-1)^{3(B-L)}$ are the spin and 
matter parity, respectively. Here B and L stand for Baryon and Lepton number. The possible implications 
of the conservation or violation of this discrete symmetry have been studied quite intensively in the last 30 years 
by many experts. See for example Refs.~\cite{Aulakh:1982yn, Hayashi:1984rd, Krauss:1988zc,Barbier:2004ez,Nath:2006ut}. 
However, there are only a few phenomenological studies of theories which dynamically explain the origin of R-parity. 
Recently, we initiated such a study in Ref.~\cite{letter} and extend its scope in this article.  
 
The simplest way to understand the state of R-parity is in the context of a B-L extension of the MSSM, 
where matter parity is just a subgroup of the new abelian symmetry, $U(1)_{B-L}$. 
These theories are quite simple because only three copies of right-handed neutrinos are needed 
for an anomaly free theory. Recently, it was noticed that the minimal B-L model violates R-parity~\cite{PRL}, 
a scenario further motivated by string theory~\cite{Braun:2005nv}. However, since only 
experiments will reveal the validity of this symmetry, it is important to understand the second 
possibility as well, \textit{i.e.} the dynamical conservation of R-parity. 
This is especially crucial because observation of missing energy signals at the LHC do not necessarily bare 
cosmological significance. Therefore, observing both missing energy and the signals discussed in this 
paper could increase the connection of dark matter to missing energy.

In the simplest framework for dynamical R-parity conservation, B-L is broken at the TeV scale 
making the model testable at the LHC. We discuss the prospects for testing the mechanism 
for the stability of the LSP in two different models which fit in this framework. Our key 
findings center around the properties of the B-L Higgs which can decay into two 
right-handed neutrinos in the first model and into two sfermions in the 
second case. The final states in the former case are especially interesting since even though 
R-parity is conserved, the final states can violate lepton number. Furthermore 
the right-handed neutrinos are long-lived giving rise to up to four displaced vertices. 
The main production channels for the Higgses at the LHC are investigated in detail 
and we discuss all possible signals one could use to test the theory of R-parity conservation. 

This work is organized as follows: In Section II we briefly summarize the main implications from R-parity conservation 
or violation. The simplest theoretical frameworks for R-parity conservation are discussed in Section III.
In Section IV we discuss the decays of the $Z_{BL}$ gauge boson including the effects of supersymmetric particles. All production 
mechanisms at the LHC for the B-L Higgses are investigated in Section V. The decays of the physical 
Higgses are discussed in Section VI, while in Section VII we study the most generic signals coming from R-parity 
conservation. Finally, we summarize our results in Section VIII.


\section{Supersymmetry, R-Parity and the LHC}
The signals indicating a discovery of low scale supersymmetry (SUSY) at 
the LHC depend on the conservation or violation of R-parity. In fact, both 
the cosmological and phenomenological aspects of the MSSM crucially depend on this. 
It is well-known that one has the following predictions: 
\begin{itemize}
\item \textit{R-Parity Conservation}: SUSY particles are produced in pairs and typically 
decay via long decay chains with multijets, multileptons and missing energy. The latter is due to the 
LSP, which is stable. Detecting missing energy is then a direct evidence for SUSY dark matter. However, 
while the LSP may be stable on collider scales, its stability on cosmological scales 
is not assured. If the mechanism for the LSP stability (R-parity conservation) 
is also tested it can shed further light on this issue~\cite{letter}.
\item \textit{R-Parity Breaking}: One can have single production of supersymmetric 
particles and possible observation of lepton and/or baryon violation at the LHC. See Ref.~\cite{Barbier:2004ez} 
for a review and Ref.~\cite{DeCampos:2010yu,Dreiner:2011xa,Bandyopadhyay:2010cu,Mukhopadhyaya:2010qf} 
for recent studies. Lepton number violation stems from non-vanishing couplings of the 
type $LH_u$, $LLe^c$ or $QLd^c$, while the presence of $u^c d^c d^c$ lead to baryon 
number violation. However, the presence of both lepton and baryon number violating 
terms together would lead to catastrophic proton decay~\cite{Nath:2006ut}.

In general it is easier to discover SUSY at the LHC if R-parity is broken since SUSY 
particles decay to SM final states instead of missing energy, except for the SM neutrinos. 
In models with spontaneous R-parity breaking through the vacuum expectation value 
of the right-handed sneutrinos~\cite{Hayashi:1984rd, Mohapatra:1986aw, Masiero:1990uj, PRL}, 
only the bilinear term $LH_u$ from above exists at the renormalizable level. Furthermore, 
it is important to note that even when R-parity is broken, the gravitino can still be a good 
dark matter candidate~\cite{Takayama:2000uz}. We postpone discussing the 
LHC testability of the theories with spontaneously broken R-parity to a later article.

\end{itemize}

If SUSY is discovered at the LHC with missing energy, a possible next step is to test the mechanism 
responsible for R-parity conservation. In the simplest case of a gauged B-L symmetry, which we will pursue here, 
the following items should be searched for:

\begin{itemize}

\item The new neutral gauge boson, $Z_{BL}$, associated with the local B-L symmetry.  
For a review on $Z^{'}$ gauge bosons see Ref.~\cite{Langacker}. See also Ref.~\cite{Basso:2010pe}.

\item The right handed neutrinos necessary for an anomaly free gauged B-L theory and study 
their decays. One possibility is through the production mechanism, 
$pp \to Z_{BL}^* \to NN$.   See for example Ref.~\cite{Huitu:2008gf,AguilarSaavedra:2009ik,TypeI-Zprime} 
for a detailed study.

\item Identify the properties of the Higgses responsible for breaking B-L.  As will be discussed later, 
these have different relationships to the different LSPs (potential dark matter candidates) and so 
studying their properties may also help to identify the dark matter candidate.

\end{itemize}

There are several studies on the discovery of the first two points: $Z^{'}$ gauge bosons and right-handed neutrinos at the LHC. 
However,  the properties of the SUSY Higgs bosons responsible for the conservation of R-parity have not been studied, 
except in Ref.~\cite{letter}, which expand upon here by studying the Higgs production and decay in more detail.


\section{Theoretical Framework for R-Parity Conservation}
The simple B-L extension of the MSSM has two different incarnations which carry a mechanism for dynamically conserving R-parity. 
Before addressing these, we briefly review the status of R-parity in the MSSM.

As it is well-known, the superpotential of the MSSM is given by
\begin{equation}
{\cal W}_{MSSM}= {\cal W}_{RpC} \ + \ {\cal W}_{RpV},
\end{equation}
where ${\cal W}_{RpC}$ is the R-parity conserving part 
\begin{eqnarray}
	{\cal W}_{RpC} &= & \ Y_u \ \hat{Q} \ \hat{H}_u \ \hat{u}^c 
		\ + \ Y_d \ \hat{Q} \ \hat{H}_d \  \hat{d}^c
		\ + \ Y_e \ \hat{L} \  \hat{H}_d \ \hat{e}^c
		\ + \ \mu \ \hat{H}_u \ \hat{H}_d,
\end{eqnarray}
and 
\begin{equation}
\label{RPV}
{\cal W}_{RpV} = \epsilon \hat{L} \ \hat{H}_u \ + \  \lambda \hat{L} \ \hat{L} \  \hat{e}^c \ + \  
\lambda' \hat{Q} \ \hat{L} \ \hat{d}^c \ + \  \lambda'' \hat{u}^c \  \hat{d}^c \  \hat{d}^c,  
\end{equation}
contains the R-parity violating terms. Gauging B-L forbids the terms in Eq.~(\ref{RPV}), which all violate B-L by one unit. 
The most straightforward possibility for the new gauge group is
\begin{equation}
	\label{gauge.group}
	SU(3)_C \bigotimes SU(2)_L \bigotimes U(1)_Y \bigotimes U(1)_{B-L}
\end{equation}
Since three copies of right-handed neutrinos are needed to cancel linear and cubic B-L anomalies, the most general superpotential becomes
\begin{equation}
{\cal W}_{B-L}= {\cal W}_{RpC} \ + \ Y_\nu \hat{L} \hat{H}_u \hat{\nu}^c \ + \ {\cal W}_{extra}, 
\end{equation}
where the last term is model dependent. The particle content and its charge under Eq.~(\ref{gauge.group}) is that of the MSSM:
\begin{align}
\begin{split}
	\hat{Q}^T &= \left(\hat{u}, \hat{d}\right) \sim (3,2,1/6,1/3),
	\hspace{1.3 cm}
	\hat{u}^c \sim (\bar{3},1,-2/3,-1/3),
	\hspace{1 cm}
	\hat{d}^c \sim 	(\bar{3}, 1, 1/3,-1/3),
\\
	\hat{L}^T &= \left(\hat{\nu}, \hat{e}\right) \sim (1,2,-1/2, -1),
	\hspace{1.3 cm}
	\hat{e}^c \sim (1, 1, 1, 1),
\\
	\hat{H}_u^T &= \left(\hat{H}_u^+, \hat{H}_u^0 \right) \sim (1,2,1/2,0),
	\hspace{.8cm}
	\hat{H}_d^T = \left(\hat{H}_d^0, \hat{H}_d^- \right) \sim (1,2, -1/2,0),
\end{split}
\end{align}
plus the right-handed neutrinos:
\begin{equation}
	\hat{\nu}^c \sim (1,1,0,1).
\end{equation}
The only remaining sector left to specify is the Higgs content which serves to break $U(1)_{B-L}$ and also governs the dynamical conservation of 
R-parity.  Here we will introduce two possibilities within the simple framework of adding only a vector-like pair of Higgses.  In general, we will refer to these Higgses as
\begin{align}
	\hat {\bar \phi} \sim (1,1,0, \eta_\phi) \quad \quad
	\hat \phi \sim (1,1,0, -\eta_\phi).
\end{align}

\begin{itemize}
\item {\bf Model I ($\mathbf{\eta_\phi = 2}$)}:  Here we dub the Higgses $\hat{X}, \hat{\bar{X}} \sim (1,1,0,\pm 2)$. The extra term in the above superpotential reads as:
\begin{equation}
{\cal W}_{extra}^{(I)}=\mu_X \hat{X} \hat{\bar{X}} \ + \ f  \hat{\nu}^c \hat{\nu}^c \hat{X}.
\end{equation}
Once the Higgses acquire a VEV, the second term above induces a Majorana mass term for the right-handed neutrinos making the neutrinos Majorana fermions.  Furthermore, the new Higgses can decay at tree level into two right-handed neutrinos. Recently, it was noted that radiative symmetry 
breaking via the $f$ Yukawa coupling dictates that in the majority of the parameter space R-parity is spontaneously broken~\cite{Fate}. 
In this paper, we do not subscribe to any high-scale scenario and simply assume that the Higgses acquire an R-parity conserving VEV and then study their signals at the LHC. Interestingly enough, even though R-parity is conserved, lepton number is still broken and could manifest itself 
in the form of same-sign leptonic final states.  For lepton flavor violating rare leptonic decays, see~\cite{Babu:2002tb}. For the study of other aspects of this model see Ref.~\cite{Krauss:1988zc,Huitu:2008gf}.
\item {\bf Model II ($\eta_\phi = \frac{2p}{2q+1}$)}: While it is well-known that Higgs bosons with even B-L charge which acquire a VEV conserve R-parity~\cite{Krauss:1988zc}, we supplement this by noting that $2p/(2q+1)$ with $p$ and $q$ integers also conserves R-parity.  This includes $\eta_\phi = 4, \ 2/3$ and $4/3$ for example. This model has not been studied before and has distinctly different Higgs physics from Model I.  We term the Higgses in this case $\hat{S}, \hat{\bar{S}} \sim (1,1,0, \pm \eta_S)$ and the extra term in the superpotential 
is simply the mass term:
\begin{equation}
{\cal W}_{extra}^{(II)}=\mu_S \hat{S} \hat{\bar{S}}.
\end{equation}
Neutrinos in this case are Dirac fermions and the new physical Higgses do not couple to the MSSM superfields 
at tree level. This scenario is quite interesting because it is so distinct from the previous case indicating 
different signatures for the mechanism responsible for the stability of the LSP and give rise to very exotic Higgs signals at the LHC.

\end{itemize}

In general models of $B-L$, such as the Models I and II, kinetic mixing is possible between the $Z$ and $Z_{BL}$.  However the mixing is constrained to be less than about $10^{-2}$ and only plays a role in precision physics~\cite{Rizzo:2006nw}.  We therefore ignore it for the remainder of this work.

For the remainder of this section we discuss the details of these two scenarios in a general way.

\subsection{B-L Symmetry Breaking} 
In order to discuss the symmetry breaking in these models in a general way, we use the notation 
$\phi,\bar{\phi} \sim (1,1,0,\pm n_{\phi})$. Then, $\phi(\bar{\phi})$ can be $X(\bar{X})$ 
in model I or $S(\bar{S})$ in model II. The relevant soft terms for our discussions are:
\begin{eqnarray}
- {\cal L}_{Soft} &\supset & \left( a_\nu \tilde{L} H_u \tilde{\nu}^c \ - \ b_{\phi} \phi \bar{\phi} 
\ + \  \frac{1}{2} M_{BL} \tilde{B}' \tilde{B}'  \ + \  \text{h.c.} \right) 
\nonumber \\
& + &  m_\phi^2 |\phi|^2 \ + \ m_{\bar{\phi}}^2 |\bar{\phi}|^2 \ + \ m_{\tilde{\nu}^c}^2 |\tilde{\nu}^c|^2 + ...,
\end{eqnarray}
where $\tilde{B}'$ is the B-L gaugino and $...$ indicates MSSM soft terms. Spontaneous B-L breaking and R-parity conservation require
the nonzero VEVs for $\phi$ and $\bar{\phi}$.  Notice that in the above equation one should add the trilinear term $a_f \tilde{\nu}^c \tilde{\nu}^c \phi$ in the case of Model I.
Using $\left<\phi \right>=v/\sqrt{2}$ and $\left<\bar{\phi}\right>=\bar{v}/\sqrt{2}$ one finds
\begin{eqnarray}
V&=& \frac{1}{2}|\mu_{\phi}|^2 \left( v^2 + \bar{v}^2\right) \ - \ b_{\phi} v \bar{v} \ + \ \frac{1}{2}m_{\phi}^2 v^2 \ + \ \frac{1}{2} m_{\bar{\phi}}^2 \bar{v}^2 
 \ + \  \frac{g_{BL}^2}{32} n_{\phi}^2 \left( v^2 - \bar{v}^2 \right)^2.
\end{eqnarray}
This form is very similar to that of the MSSM and the derivations that follow mirror those of the MSSM with the appropriate replacements. 
Assuming that the potential is bounded from bellow along the D-flat direction leads to the condition:
\begin{equation}
2 b_{\phi} < 2 |\mu_{\phi}|^2 + m_{\phi}^2 + m_{\bar{\phi}}^2,
\end{equation} 
while 
\begin{equation}
b_{\phi}^2 > \left( |\mu_{\phi}|^2 + m_{\phi}^2 \right) \left( |\mu_{\phi}|^2 + m_{\bar{\phi}}^2 \right).
\end{equation}
is necessary for a nontrivial minimum. Minimizing with respect $v$ and $\bar{v}$ one gets
\begin{eqnarray}
\begin{split}
|\mu_{\phi}|^2 + m_{\phi}^2 - \frac{1}{2} m_{Z_{BL}}^2 \cos 2 \beta' - b_{\phi} \cot \beta'=0,
\\
|\mu_{\phi}|^2 + m_{\bar{\phi}}^2 + \frac{1}{2} m_{Z_{BL}}^2 \cos 2 \beta' - b_{\phi} \tan \beta'=0,
\end{split}
\end{eqnarray}
with $\tan \beta'=v/\bar{v}$ and $m_{Z_{BL}}^2=g_{BL}^2 n_{\phi}^2 (v^2 + \bar{v}^2)/4$.  These can be recast into the more useful form:
\begin{align}
	\frac{1}{2} m_{Z_{BL}}^2 & = -\left| \mu_{\phi} \right|^2 - \left(
		\frac{m_\phi^2 \tan^2 \beta' \ - m_{\bar \phi}^2}{\tan^2 \beta' - 1}\right),
	\\
	b_\phi & = \frac{\sin 2 \beta^{'}}{2} \left(2\left| \mu_{\phi} \right|^2 + m_\phi^2 + m_{\bar \phi}^2 \right).
\end{align}
From here we move on to describe the spectrum details.
\subsection{Mass Spectrum}

\underline{Higgs Bosons}: 

The physical Higgs content includes the MSSM Higgses: $h$, $H$, $A$, $H^{\pm}$, as well as two extra CP-even neutral 
Higgses, $H_1$ and $H_2$, and one CP-odd Higgs, $A_{\phi}$ ($X_1$, $X_2$ and $A_{BL}$ in Model I and $S_1$, $S_2$ and $A_S$ in Model II).  The complex gauge states can be written down in terms of their real components:
\begin{equation}
\phi=\frac{1}{\sqrt{2}}(v + \phi_R) + \frac{i}{\sqrt{2}} \phi_I,  \hspace{2.0cm} 
\bar{\phi}=\frac{1}{\sqrt{2}}( \bar{v} + \bar{\phi}_R) + \frac{i}{\sqrt{2}} \bar{\phi}_I,
\end{equation}
and related to the physical states through
\begin{align}
\label{GaugeToMass}
\begin{pmatrix}
	\phi_R
	\\
	\bar{\phi}_R
\end{pmatrix}
	& =
\begin{pmatrix}
	\cos \alpha'
	&
	\sin \alpha'
	\\
	- \sin \alpha'
	&
	\cos \alpha'
\end{pmatrix}
\begin{pmatrix}
	H_1
	\\
	H_2
\end{pmatrix},
\\
\begin{pmatrix}
	\phi_I
	\\
	\bar{\phi}_I
\end{pmatrix}
	& =
\begin{pmatrix}
	\sin \beta'
	&
	\cos \beta'
	\\
	- \cos \beta'
	&
	\sin \beta'
\end{pmatrix}
\begin{pmatrix}
	G_\phi
	\\
	A_\phi
\end{pmatrix},
\end{align}
where $G_\phi$ is the Goldstone boson associated with breaking B-L and which is eaten by $Z_{BL}$. 
The Higgs spectrum is completely parameterized by: $\tan \beta', m_{Z_{BL}}$ and 
\begin{equation}
m_{A_\phi} ^2= \frac{2 b_\phi}{\sin 2 \beta^{'}} . 
\end{equation}
The eigenvalues and the mixing angles in the CP-even neutral Higgs sector read as
\begin{widetext}
\begin{equation}
	m_{H_{1,2}}^2=\frac{1}{2} \left( m_{A_\phi}^2 + m_{Z_{BL}}^2 
		\mp \sqrt{(m_{A_\phi}^2 - m_{Z_{BL}}^2)^2  +  4 m_{Z_{BL}}^2 m_{A_\phi}^2 \sin^2 (2 \beta')  } \right),
\end{equation}
\end{widetext}
\begin{equation}
\frac{\tan 2 \alpha'}{\tan 2 \beta'} = \frac{m_{A_\phi}^2 +  m_{Z_{BL}}^2}{m_{A_\phi}^2 -  m_{Z_{BL}}^2}.
\end{equation}
Notice that in the limit, $m_{Z_{BL}}^2 \gg m_{A_\phi}^2$, which will be employed later, the above simplifies to
\begin{eqnarray}
	m_{H_1}^2 &\sim& m_{A_\phi}^2 \left(  1 - \sin^2 2 \beta' \right), \\
	m_{H_2}^2 &\sim & m_{Z_{BL}}^2 + m_{A_\phi}^2 \sin^2 2 \beta', \\
	\alpha' &\sim & - \beta' - \frac{\tan 2 \beta'}{1+ \tan^2 2 \beta'} \frac{m_{A_\phi}^2}{m_{Z_{BL}}^2}.
\end{eqnarray}
Then, assuming a TeV scale $m_{Z_{BL}}$ and small $m_{A_\phi}$ one expects two light Higgses at around the same mass: $H_1$ and $A_{\phi}$, and a heavy one, $H_2$ close to the $Z_{BL}$ mass.  Regardless of the parameter space though, the following relationships are observed: $m_{H_1} \leq m_{A_{\phi}} \text{ and } m_{Z_{BL}}$ and $m_{H_2} \geq m_{A_{\phi}} \text{ and } m_{Z_{BL}}$.

\underline{Neutrino Sector}:

The neutrino sector of the two models differs dramatically so we can not discuss the two models generically in a worthwhile way.  Model II is simple, in this case, neutrinos are Dirac fermions.  However, in Model I, once the $X$ and $\bar X$ get a VEV a Majorana mass term will be induced for the right-handed neutrinos:
\begin{equation}
	m_{N_i} = \sqrt 2 \, f_i \sin \beta' \, \frac{m_{Z_{BL}}}{g_{BL}},
\end{equation}
noting that $f$ can be diagonalized without loss of generality. This mass in turn triggers the type I seesaw mechanism for neutrino masses~\cite{TypeI}:
\begin{equation}
	m_\nu = \sqrt 2 \, v_u^2 Y_\nu^T (m_N)^{-1} Y_\nu.
\end{equation}
As is typical for TeV scale seesaws, $Y_\nu \sim 10^{-6}$ correctly reproduces the neutrino masses.

\underline{Neutralino Sector}: 

The neutralino mass matrix, in the basis $\tilde \psi^0 = \left(\psi_\text{MSSM}, \, \tilde B', \, \tilde \phi, \, \tilde{\bar \phi} \right)$, reads as:
\begin{equation}
	\mathcal{M}_{\chi^0} = \left(
	\begin{array}{cccc}
		{\cal M}_\text{MSSM}
		&
		0
		&
		0
		&
		0
		\\
		0
		&
		M_{BL}
		&
		-g_{BL} \  \frac{n_{\phi}}{2} \ v 
		&
		g_{BL} \  \frac{n_{\phi}}{2} \  \bar v
	\\
		0
		&
		-g_{BL} \  \frac{n_{\phi}}{2} \ v 
		&
		0
		&
		-\mu_\phi
	\\
		0
		&
		g_{BL} \  \frac{n_{\phi}}{2} \  \bar v 
		&
		-\mu_\phi
		&
		0
	\end{array} \right),
\end{equation}
where $\psi_\text{SM}$ defines the MSSM neutralinos and ${\cal M}_\text{MSSM}$ the 
four-by-four MSSM neutralino mass matrix. Here, we define the mass eigenstates as in the MSSM
\begin{equation}
\label{neutralino.mixing}
	\tilde \chi^0_i = N_{ij} \tilde \psi_j^0,
\end{equation}
where $N$ diagonalizes the full seven-by-seven neutralino mass matrix and breaks up into block 
diagonal form where the upper  four-by-four block diagonalizes the MSSM and the lower 
three-by-three block diagonalizes the B-L neutralino sector.  The eigenstates are labeled 
with increasing mass so that $\chi^0_1 \ (\chi^0_7)$ is the lightest (heaviest) neutralino, although 
the lightest B-L neutralino will play a role later so we denote it $\tilde \chi_{BL}$.

\underline{Sfermion Masses}:

In the sfermion sector, the mass matrices $\mathcal{M}_{\tilde u}^2$, $\mathcal{M}_{\tilde d}^2$, 
and $\mathcal{M}_{\tilde e}^2$ in the basis $\left(\tilde f_L, \ \tilde f_R\right)$ are given by
\begin{eqnarray}
	&&\left(\begin{array}{cc}
		m_{\tilde Q}^2
		\ + \ m_{u}^2
		\ + \  \left(\frac{1}{2} \ - \  \frac{2}{3} s_W^2 \right) \ M_Z^2 \ c_{2\beta} 
		\ + \ \frac{1}{3} D_{BL}
		&
		\frac{1}{\sqrt 2} \left(a_u \ v_u - Y_u \ \mu \ v_d\right)
	\\
		\frac{1}{\sqrt 2} \left(a_u \ v_u - Y_u \ \mu \ v_d\right)
		&
		m_{\tilde u^c}^2
		\ + \ m_{u}^2
		\ + \ \frac{2}{3} M_Z^2 \ c_{2\beta}\ s_W^2 
		\ - \ \frac{1}{3} D_{BL}
	\end{array}\right),
	\nonumber
\end{eqnarray}	
\begin{eqnarray}	
	&& \left(\begin{array}{cc}
		m_{\tilde Q}^2
		\ + \ m_{d}^2
		\ - \  \left(\frac{1}{2}  \ - \ \frac{1}{3} \  s^2_W \right) M_Z^2 \ c_{2 \beta}
		\ + \ \frac{1}{3} D_{BL}
		&
		\frac{1}{\sqrt 2} \left(Y_d \ \mu \ v_u - a_d \ v_d\right)
	\\
		\frac{1}{\sqrt 2} \left(Y_d \ \mu \ v_u - a_d \ v_d\right)
		&
		m_{\tilde d^c}^2
		\ + \ m_{d}^2
		\ - \  \frac{1}{3} \ M_Z^2  \ c_{2\beta} \ s^2_W
		\ - \ \frac{1}{3} D_{BL}
	\end{array}\right),
	\nonumber \\
	\\
	&& \left(\begin{array}{cc}
		m_{\tilde L}^2
		\ + \ m_{e}^2
		\ - \ \left( \frac{1}{2} \ - s_W^2 \right) M_Z^2 \ c_{2\beta} 
		\ - \ D_{BL}
		&
		\frac{1}{\sqrt 2} \left(Y_e \ \mu \ v_u - a_e \ v_d\right)
	\\
		\frac{1}{\sqrt 2} \left(Y_e \ \mu \ v_u - a_e \ v_d\right)
		&
		m_{\tilde e^c}^2
		\ + \ m_{e}^2
		\ - \  M_Z^2 \ c_{2\beta}\ s_W^2
		\ + \ D_{BL}
	\end{array}\right),
	\nonumber 
\end{eqnarray}
where $c_{2\beta} = \cos 2\beta$, $s_W = \sin\theta_W$ and
\begin{equation}
D_{BL} \equiv \frac{1}{8} \ g_{BL}^2  n_{\phi} \left( \ \bar v^2 - v^2  \right)= \frac{1}{2 n_{\phi}} M_{Z_{BL}}^2 \cos 2 \beta'.
\end{equation}
$m_u, \ m_d$ and $m_e$ are the respective fermion masses and $a_u, \ a_d$ and $a_e$
are the trilinear $a$-terms corresponding to the Yukawa couplings $Y_u, \ Y_d$ and $Y_e$.
Typically, it is assumed that substantial left-right mixing occurs only in the third generation.  
Regardless, the physical states are related to the gauge states by
\begin{align}
\label{eq_squarkmixing}
	\begin{pmatrix}
		\tilde q_1
		\\
		\tilde q_2
	\end{pmatrix}
	& =
	\begin{pmatrix}
		\cos \theta_{\tilde q}
		&
		\sin \theta_{\tilde q}
		\\
		- \sin \theta_{\tilde q}
		&
		\cos \theta_{\tilde q}
	\end{pmatrix}
	\begin{pmatrix}
		\tilde q
		\\
		\tilde q^{c*}
	\end{pmatrix},
\end{align}
where here we are thinking about the squark sector, but, of course, the same thing can be done in the slepton sector.

The left-right mixing in the sneutrino sector is negligible due to the small Dirac Yukawa couplings necessary 
for the type I seesaw mechanism. The left-handed masses are
\begin{equation}
	m_{\tilde \nu_L}^2 = \ m_{\tilde L}^2
		\ + \  \frac{1}{2} M_Z^2 \cos 2 \beta
		\ - \ D_{BL}.
\end{equation}
In Model I the right-handed sneutrino CP-even and CP-odd states are split by trilinear terms involving the B-L 
Higgses. Remembering that the Yukawa matrix, $f$, can be chosen to be diagonal without loss of generality,
the masses of the right-handed sneutrinos are given by
\begin{align}
	\label{m.NI}
	m_{\tilde N_{R_i}}^2 & = m_{\tilde \nu_i^c}^2
		\ + \ 2 f_i^2 \ v^2
		\ + \ \sqrt 2 \ a_{f_i} \ v
		\ + \ \sqrt 2 \ f_i \ \mu_X \ \bar v
		\ + \ D_{BL},
	\\
	\label{m.NR}
	m_{\tilde N_{I_i}}^2 & = m_{\tilde \nu_i^c}^2
		\ + \ 2 f_i^2 \ v^2
		\ - \ \sqrt 2 \ a_{f_i} \ v
		\ - \ \sqrt 2 \ f_i \ \mu_X \ \bar v
		\ + \ D_{BL}.
\end{align}
where $i$ runs over all three generations and repeated indices are not summed. The masses for Model II can be recovered from the above by setting $f_i, \, a_{f_i} \to 0$ and in this case the right-handed sneutrino can be treated as a single complex scalar field.

It is important to reemphasize that in this context R-parity conservation and therefore the stability of the LSP is a direct consequence of the breaking of 
B-L via $\phi$ and $\bar \phi$.  The properties of these fields are the crucial ingredient for testing this mechanism.  These can give rise to unique signals: lepton number violating in Model I (despite the conservation of R-parity) and multi-leptons and/or multi-photons in Model II.


\section{Decays of the $\ZBL$ Neutral Gauge Boson}
The discovery of a new B-L gauge boson at the LHC is crucial to establish the existence 
of a new abelian gauge symmetry and to test the mechanism responsible for R-parity conservation 
or violation in the supersymmetric case. In this section we discuss the main features of the 
$Z_{BL}$ boson decays in order to understand the impact of the supersymmetric particles on the 
total width. Furthermore, $Z_{BL}$ can decay into the Higgses $A_{BL}$ and $X_1$, a 
decay that does not exist in the minimal non-SUSY B-L model (since there is no $A_{BL}$). 
As we will discuss later, this decay width also enters into the Higgs pair production cross section, 
$pp \to Z^*_{BL} \to H_1 A_{\phi}$, which is an important channel for the discovery of these fields.

The current bounds on $\ZBL$ from LEP II are commonly quoted as $m_{\ZBL}/g_{BL} > 6$ TeV
~\cite{Carena:2004xs} but since our covariant derivative is defined as 
$D^\mu = \partial^\mu - \frac{1}{2} \, g_{BL} \, n_\phi \, Z^\mu_{BL}$, the relevant bound here is
\begin{equation}
	\frac{m_{Z_{BL}}}{g_{BL}} > 3 \text{ TeV}.
\end{equation}
In what follows, we will simply take this upper limit as an equality. 

\begin{figure}[t] 
\plot{.7\linewidth}{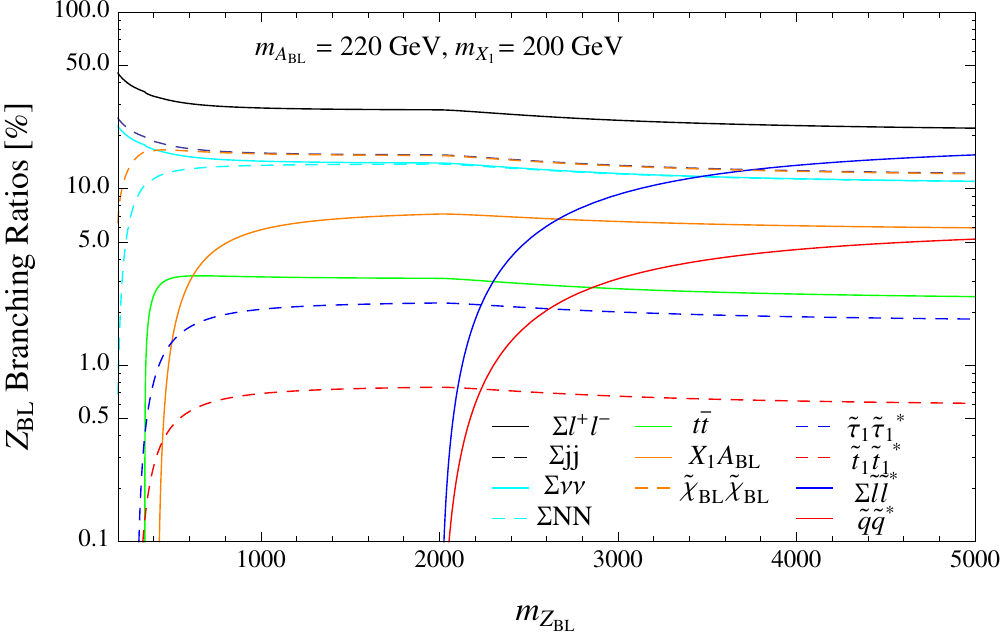}
\caption{Branching ratios of the $Z_{BL}$ boson for 
$\mabl =220$\,GeV and $\mxone=200$\,GeV.
The masses of the three right-handed neutrinos are $m_{N_i} = 95$\,GeV.
We consider one light squark, $m_{\tilde{t}_1}=150$\,GeV, and one light slepton,
$m_{\tilde{\tau}_1} = 150$\,GeV, all other eleven squarks, five charged and six neutral 
sleptons are heavy (including the three right-handed sneutrinos),
$m_{\tilde{q}} = m_{\tilde{l}} = 1$\,TeV. All mixing angles in the sfermion sector 
are set to zero. The neutralino masses are determined by $\mu_X$ and $M_{BL}$, 
both taken here to be $150$ GeV.}
\label{fig_Zbranchingratios}
\end{figure}

The $\ZBL$ boson can decay into a pair of fermions, light or heavy neutrinos, 
sfermions, or into a pair of two new Higgs boson. The partial widths for the decay into particles 
$P_1, P_2$ of masses $m_1, m_2$ are given by,
\begin{align}
	\Gamma(\ZBL \to P_1 P_2) = 
	\frac{1}{16 \pi \mzbl} \, \left| \overline{\mm} (\ZBL \to P_1 P_2) \right|^2 \, 
	\sqrt{ \left( 1 - \frac{(m_1 +m_2)^2}{\mzbl^2} \right) 
		 \left( 1 - \frac{(m_1 -m_2)^2}{\mzbl^2} \right)
	},
\end{align}
where the squared matrix elements follow from the Feynman rules in the Appendix:
\begin{align}
\label{Zff}
 \left| \overline{\mm} (\ZBL \to f_i \bar{f_i}) \right|^2 =&~
	\frac{4}{3} c_{f} \left( \frac{g_{BL} }{2}n_{BL}^{f}\right)^2 \mzbl^2 
		\left(1+\frac{2 m_{f_i}^2}{\mzbl^2} \right),
	 \qquad f_i = u,d,c,s,b, t, e, \mu, \tau;	
\\
 \left| \overline{\mm} (\ZBL \to \nu_i \bar{\nu_i}) \right|^2 =&~
	 \frac{2}{3} \left( \frac{g_{BL} }{2}n_{BL}^{\nu}\right)^2 \mzbl^2,
\\
 \left| \overline{\mm} (\ZBL \to  \bar{N} N) \right|^2 =&~
	 \frac{2}{3} \left( \frac{g_{BL} }{2}n_{BL}^{\nu_R}\right)^2
	\, \mzbl^2 \,  
	\left(1 - 4\frac{m_{N}^2}{\mzbl^2}\right),
\end{align}	
\\	
\begin{align}
 \left| \overline{\mm} (\ZBL \to \tilde{f}_{\alpha} \tilde{f}^\ast_{\beta}) \right|^2  =&~
	\frac{1}{3} c_{\tilde{f}}
	\left( \frac{g_{BL} }{2}n_{BL}^{\tilde{f}}\right)^2 \mzbl^2
	\left( 1 - \frac{2 m_{\tilde{f}_{\alpha}}^2 + 2 m_{\tilde{f}_{\beta}}^2}{\mzbl^2}
		+   \frac{\left(m_{\tilde{f}_{\alpha}}^2 - m_{\tilde{f}_{\beta}}^2\right)^2}{\mzbl^4}
	\right)
\\&
	\times~
	\left( U_{\alpha 1}^{\tilde{f}} U_{\beta 1}^{\tilde{f}} + 
		U_{\alpha 2}^{\tilde{f}} U_{\beta 2}^{\tilde{f}} \right)^2,
	\qquad\quad \tilde{f}_{\alpha} \tilde{f}_{\beta}^* = 
		\tilde{q}_{i\alpha} \tilde{q}_{i\beta}^{*},
		\tilde{l}_{i\alpha} \tilde{l}_{i\beta}^{*},
		\tilde{\nu}_i \tilde{\nu}_{i}^{*},
		\tilde{\nu}_{Ri} \tilde{\nu}_{Ri}^{*};
\nn
\\
 \left| \overline{\mm} (\ZBL \to X_i \ABL) \right|^2  =&~
	\frac{1}{3} \left( \frac{g_{BL} }{2}n_{BL}^{X}\right)^2 \mzbl^2
	\left( 1 - \frac{2 m_{X_{i}}^2 + 2 \mabl^2}{\mzbl^2}
		+   \frac{\left(m_{X_i}^2 - \mabl^2\right)^2}{\mzbl^4}
	\right)
\\&
	\times~ \cos^2(\beta'-\alpha'),
\nn
\\
 \left| \overline{\mm} (\ZBL \to \bar \chi_i \chi_j) \right|^2  =&~
 	\frac{1}{3} \left( \frac{g_{BL} }{2}n_{BL}^{X}\right)^2
	\mzbl^2
	\left(
		1 - \frac{m_i^2 + m_j^2 + 6 m_i m_j}{2 \mzbl^2} - \frac{\left(m_i^2 - m_j^2\right)^2}{2 \mzbl^4}
	\right)
\\&
	\notag
	\times~
	\left(
		N_{i \tilde{\bar X}} N^\dagger_{\tilde{\bar X} j} - N_{i \tilde X} N^\dagger_{\tilde X j}
	\right)^2 \left(1+\delta_{ij}\right).
\end{align}
Here, $i$ is a generation index, $c_f$ are color factors ($c_{q_i} = 3$, $c_{l_i}=1$) and $U^{\tilde{f}}$ 
are the unitary sfermion mixing matrices introduced in Eq.~\eqref{eq_squarkmixing} and $N_{ij}$ are the neutralino mixing matrices defined in Eq.~(\ref{neutralino.mixing}).   Using the above expressions we show the branching ratios of $Z_{BL}$ in Fig.~\ref{fig_Zbranchingratios}. 
In order to simplify our analysis, we choose the masses of the three right-handed neutrinos, $m_{N_i} = 95$\,GeV. 
We consider one light squark, $m_{\tilde{t}_1}=150$\,GeV, and one light slepton, $m_{\tilde{\tau}_1} = 150$\,GeV.  
All other eleven squarks, five charged and six neutral sleptons are heavy (including the three right-handed sneutrino),
$m_{\tilde{q}} = m_{\tilde{l}} = 1$\,TeV. All mixing angles in the sfermion sector are set to zero for simplicity.  
The masses of the new neutralinos are determined by $\mu_X$ and $M_{BL}$, both are taken here to be $150$ GeV.  
Only the lightest state contributes, while the heavier ones have masses very close to $m_{Z_{BL}}$ and give 
negligible or zero contributions. Notice that the numerical results are shown for the model I, where the 
Higgses breaking B-L have $n_{\phi}=\pm 2$. 

Fig.~\ref{fig_Zbranchingratios} shows that once the $Z_{BL}$ mass is above 2 TeV, the ``susy threshold", 
the decays into superpartners can become important. In the scenario considered in Fig.~\ref{fig_Zbranchingratios}, for $m_{\ZBL} = 3000$ GeV we have the following approximate leading branching ratios:
\begin{itemize}
	\item $\sum l^+ l^- \sim 24.4 \%$
	\item $\sum jj \sim 13.6 \%$
	\item $\sum \nu \nu, \ \sum N N \sim 12.2 \%$
	\item SUSY $\sim 28.5 \%$
\end{itemize}
\begin{figure}[t] 
\plot{.7\linewidth}{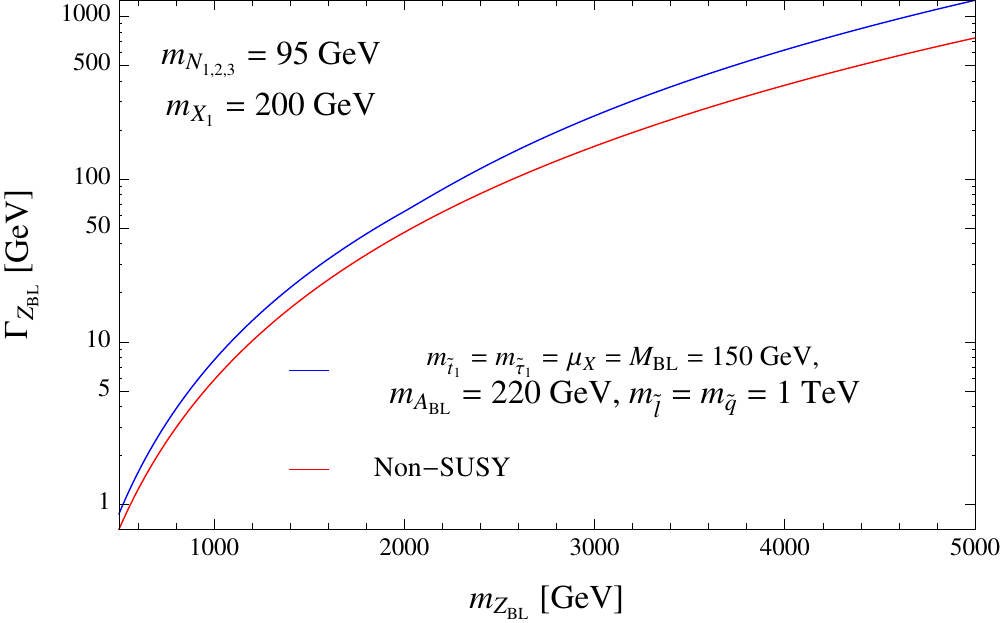}
\caption{The total $Z_{BL}$ decay width as a function of the $Z_{BL}$ mass for a a SUSY spectrum with an $A_{BL}$ mass of $220$ GeV, the lightest stop and lightest stau mass of $150$ GeV and all other sfermions at $1$ TeV in blue and $\mu_X = M_{BL} = 150$ GeV determine the neutralino masses.  In red, for comparison, is the total width for the non-SUSY case.  Note that while the decay to $X_1$ and $A_{BL}$ is not a SUSY decay, it does not exist in the minimal non-SUSY B-L model.  In both cases, all three right-handed neutrinos are assumed to be degenerate with a mass of $95$ GeV and $m_{X_1}=200$ GeV.}
\label{Z.width}
\end{figure}
The total decay width of the $Z_{BL}$ is shown in Fig.~\ref{Z.width} assuming all three right-handed 
neutrinos are degenerate with a mass of $95$ GeV and the maximum value of $g_{BL}$ consistent with 
LEP II. In order to further investigate the impact of the supersymmetric particles on the total decay width, 
we compare the total width for a given SUSY spectrum (blue line) with the non-SUSY case (red line).
For $Z_{BL}$ masses above the SUSY spectrum, the decays into supersymmetric particles contribute 
significantly and the decay widths can have significant variation between the two cases.

The key difference in the above for Model II is the different value of $n_\phi$ and the Dirac nature of the neutrinos.  In this case, the $\ZBL$ decay width to Dirac neutrinos is simply given by Eq.~(\ref{Zff}).  However, the main features of the supersymmetric contribution to the branching rations and total width are similar. 


\section{Production Mechanisms of the B-L Higgs Bosons}
\label{sec_HiggsProduction}
%
\begin{figure}[t]
\centering
\Fdiag{}{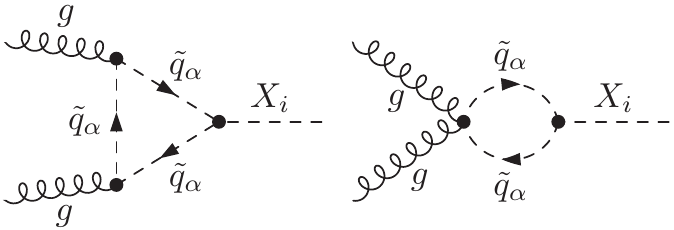}\qquad\Fdiag{}{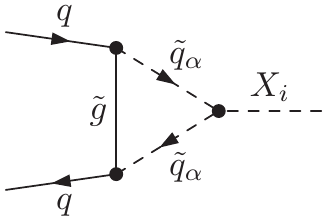} \\
(a) \\
\Fdiag{}{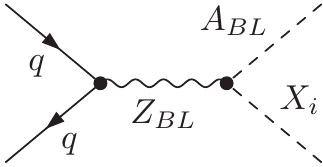}\qquad
\Fdiag{}{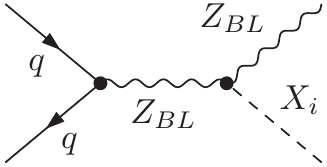}\qquad
\Fdiag{}{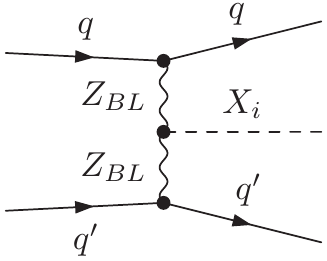}\\
(b) \hspace*{3.4cm} (c) \hspace*{3.4cm} (d)
\caption{Parton-level Feynman diagrams for the production of the CP-even Higgs $X_i$ at lowest order, via (a) single production (gluon--gluon fusion or $q\bar{q}$ annihilation), (b) $\ABL X_i$ pair production, (c) associated $\ZBL X_i$ production, 
and (d) $\ZBL$ boson fusion. For (d), 
the diagram with crossed external lines is not shown explicitly. 
The initial state particles $q,q'$ can be any of the light-flavor quarks or antiquarks.}
\label{fig_production}
\end{figure}
The dominant contributions to B-L Higgs production arise from the single CP-even production via gluon fusion 
and pair production of the CP-even and CP-odd Higgses.  Subdominant channels are  
associated $X_i  Z_{BL}$ production and $\ZBL$ boson fusion. 
The corresponding parton-level Feynman diagrams are shown in Fig.~\ref{fig_production}.

In the following we focus on the production of the lighter of the B-L Higgs bosons, $X_1$, 
in Model I. The production cross sections for the heavier Higgs boson, $X_2$, follow in complete analogy by replacing the corresponding couplings, but they are suppressed by the heavy mass and thus do not play an important role for our phenomenological studies.
The results for Model II follow by scaling with the corresponding B-L factor.
\subsection{Single Production via Gluon Fusion}
Single production is possible at the one-loop level via gluon-gluon fusion, 
$gg \to X_1$,  where squarks run inside the loop, see Fig.~\ref{fig_production}\,(a) and $q \bar{q} \to X_1$, at one-loop 
level mediated by a gluino--squark loop, see the last graph in Fig.~\ref{fig_production}, but this contribution is highly suppressed by the light quark masses and we neglect it. Both these channels depend on SUSY interactions of gauge coupling strengths between the Higgs and squarks: the $D$-terms.

The cross sections for the single production can be given in analogue 
to Higgs boson production within the MSSM~\cite{ggHiggsSUSY, qqHiggsSUSY}, making sure to include only the relevant diagrams from Fig.~\ref{fig_production}. The cross section is related to 
the decay width of the scalar and at partonic level it is given by
\begin{align}
\begin{split}
	\hat{\sigma}_{gg \to X_1} &= \frac{\pi^2}{8 \mxone^3} \, \Gamma_{X_1 \to gg}
	\, \delta\Big(1-\frac{\mxone^2}{\hat{s}}\Big),
\label{eq_prodCS_gen}
\end{split}
\end{align}
where $\hat{s}$ is the partonic center-of-mass energy squared. The decay width can be written as
\begin{align}
\begin{split}
\label{Gamma_X1_gluon}
	\Gamma_{X_1 \to gg} &= \frac{\alpha_s^2 \,\mxone^3}{128 \pi^3}\, \left|
		\sum_{\squarka} g_{\squarka}^{X_1} \, 
		\, \frac{1}{m_{\squarka}^2}\, A(\tau_{\squarka}) 
		\right|^2,
\end{split}
\end{align}
with $\tau_{\squark} = 4m_{\squark}^2/\mxone^2$ in terms of  the kinematic function
$A(\tau) = -\frac{1}{2}\, \tau \left( 1-\tau f(\tau) \right)$,
and
\begin{align}
	f(\tau) = \begin{cases}
	\arcsin^2 \left(\frac{1}{\sqrt{\tau}}\right),
		&\tau \geq 1 \\
	-\frac{1}{4} \left( \log \frac{1+\sqrt{1-\tau}}{1-\sqrt{1-\tau}} - i\pi\right)^2,
		&\tau < 1.
	\end{cases} 
\end{align}
The sum in Eq.(\ref{eq_prodCS_gen}) runs over all twelve squark eigenstates and  
the couplings $g_{\squarka}^{X_1} \equiv g_{\squarka\squarka}^{X_1}$ are given in the appendix.
Note that only the diagonal $X\squark\squark$ couplings enter since the $g\squark\squark$ couplings 
preserve gauge and mass eigenstate of the squarks. 

The couplings for the two eigenstates of a given squark $\squarka = \squark_{1,2}$ only differ by sign. 
This allows us to rewrite Eq.(\ref{eq_prodCS_gen}) as:  
\begin{align}
\begin{split}
	\hat{\sigma}_{gg \to X_1} &= \frac{\alpha_s^2}{1024 \pi} \, \left|
		\sum_{\squark} g_{\squark_1}^{X_1} \, 
		\left(\frac{1}{m_{\squark_1}^2} A(\tau_{\squark_1}) 
			 - \frac{1}{m_{\squark_2}^2} A(\tau_{\squark_2}) 
		\right)\right|^2
		\, \delta\Big(1-\frac{\mxone^2}{\hat{s}}\Big),
\label{eq_prodCS}
\end{split}
\end{align}
where now the sum runs over the six squark flavors. From this result one can see that in the case of 
degenerate squark masses, $m_{\squark_1} = m_{\squark_2}$, the contributions cancel within each 
squark flavor, due to the opposite B-L charges of the left- and right-handed squarks.

At the hadronic level, the cross section is obtained from the partonic one by the convolution,
\begin{align}
	\sigma_{pp \to X_1}(s) =
	 \int_{\tau_0}^1 \! d\tau \, \frac{d\mathcal{L}_{gg}^{pp}}{d\tau} \,  \hat{\sigma}_{gg\to X_1}
\end{align}
with $\tau = \hat{s}/s$, $s$ being the hadronic center-of-mass energy squared, and $\tau_0 = \mxone^2/s$ 
is the production threshold. The parton luminosities are given by,
\begin{align}
	\frac{d\mathcal{L}_{ab}^{AB}}{d\tau} = \frac{1}{1+\delta_{ab}} 
	\int_{\tau}^1\! \frac{dx}{x} \, \left[
		f_{a/A}(x,\mu) \, f_{b/B}\Big(\frac{\tau}{x},\mu\Big) 
	+ 	f_{a/B}\Big(\frac{\tau}{x},\mu\Big) \, f_{b/A}(x,\mu) \right],
\end{align}
where the parton distribution functions (pdfs) $f_{a/A}(x,\mu)$ parameterize the probability
of finding a parton $a$ inside a hadron $A$ with faction $x$ of the hadron momentum at 
a factorization scale $\mu$.
\begin{figure}[t] 
\plot{.49\linewidth}{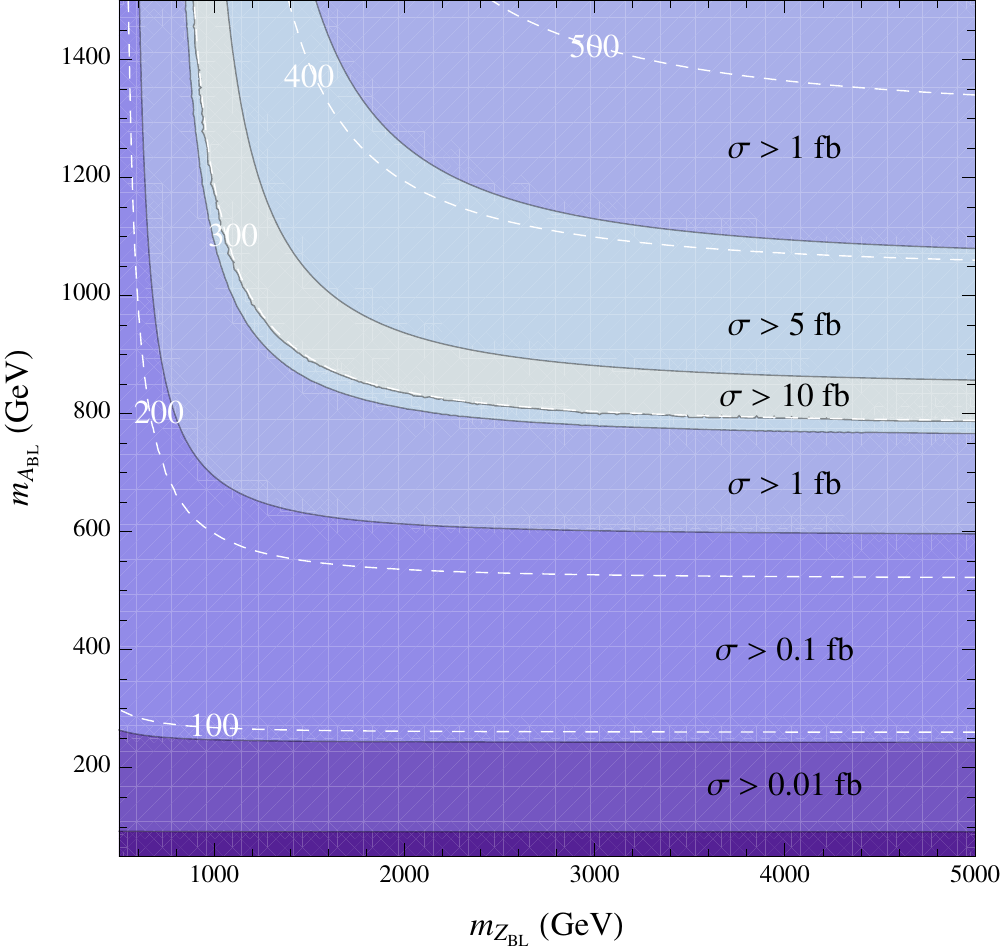}%
\plot{.5\linewidth}{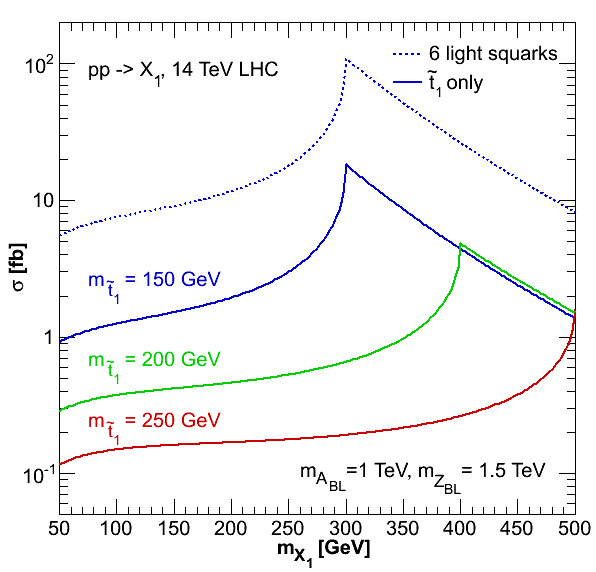}
\caption{Hadronic cross sections for single $X_1$ production at the LHC. 
One squark is considered to be light, $m_{\tilde{t}_1} = 150$\, GeV and squark mixing is neglected.
We use the MSTW 2008 LO pdf~\cite{pdf} at a central factorization scale $\mu = \mxone/2$. 
In the left panel, plot lines of constant cross section in the $\mabl-\mzbl$ with fixed $\tan\beta'=1.5$ in black and in white are lines of constant $X_1$ mass. In the right panel, the cross section is shown as a function of $\mxone$ for $\mabl = 1$\,TeV and $\mzbl = 1.5$\,TeV.  Here we also explore the possibility of six light quarks (dotted) and the effects of changing the lightest stop mass (solid).
}
\label{fig_ggXcrosssection}
\end{figure}
In Fig.~\ref{fig_ggXcrosssection} numerical predictions for the single Higgs production cross section via gluon fusion are given. The cross section depends strongly on the supersymmetric spectrum.

We consider here the conservative case where only one squark is light ($m_{\tilde{t}_1} = 150$\,GeV) 
and all other squarks are heavy and degenerate in mass and thus cancel each others contributions.
In the left panel we show the curves of constant cross section in the $\mabl-\mzbl$ plane for fixed $\tan\beta'=1.5$ in black. Lines of constant $X_1$ mass are shown in white.  The plot reflects the sharply peaked nature of the $A(\tau)$ function from Eq.~(\ref{Gamma_X1_gluon}) at $m_{X_1} = 2 m_{\tilde{t}_1}$ where the cross section can rise to about $16$ fb but then rapidly decreases when lowering the mass of $X_1$ due to the function $A(\tau)$ and when raiding the mass of $X_1$ due to both $A(\tau)$ and the decreased gluon luminosity.

In Fig.~4 (right panel) the cross section is given as a function of the Higgs mass $\mxone$, for  
fixed input parameters $\mabl=1$ TeV and $\mzbl=1.5$ TeV and one light squark, the stop (solid lines) and shows the effects of increasing the stop mass.  The cross section is very sensitive to the stop mass and decreases quickly for heavier squark masses.  To the left of the peak, the suppression is due to the function $A(\tau)$ while on peak and to the right its due to the decreased gluon luminosity.  For illustrative purposes we also consider the most optimistic case in which one squark of each flavor is light, \textit{i.e.} six light squarks (dashed line). As one can read from Eq.\eqref{eq_prodCS}, the result simply scales by six, the number of light squarks. In this case the cross section reaches $10^2$ fb.
Unfortunately, this production channel strongly depends on the SUSY spectrum and therefore 
does not allow for general predictions to test the mechanism behind R-parity conservation. An interesting property of this channel though is that a light $\ZBL$ is not necessary for production. This is different for the pair production discussed in the next subsection, 
which does not depend very strongly on the SUSY spectrum.
  
\subsection{Higgs Pair Production: {$pp \to X_1 \ABL$}} 
The Higgs pair production mechanism is the most important channel for our study and part of its interests stems from the fact that while it is not a SUSY process and is fairly independent of the SUSY spectrum (only via the $\ZBL$ width), it does not exist in minimal non-SUSY B-L models. 
The reason is of course familiar to SUSY practitioners, namely that SUSY requires vector like pairs of Higgses since these scalar fields have corresponding fermionic fields which contribute to the triangle anomalies. Therefore, a minimal non-SUSY theory has only one CP-odd scalar which becomes the longitudinal component of the $Z_{BL}$ while in SUSY there are two, one of them physical. To our knowledge and we believe for this reason, this process has not yet been discussed in the literature.  The production process proceeds via
\begin{align}
q(p_1) \, \bar{q}(p_2) \to X_1(p_3) \,\ABL(p_4).
\end{align}
The differential partonic cross sections is given by the spin- and color-averaged squared matrix element,
\begin{align}
d\hat{\sigma}_{\qqbar \to X_1\ABL} (\hat{s}) & =
	 \left| \overline{\mm}_{\qqbar \to  X_i \ABL}(\hat{s} )\right|^2 
	 \frac{d\text{PS}^{(2)}}{2\hat{s}},
\end{align}
where d$\text{PS}^{(2)}=d\hat{t}/(8\pi\hat{s})$ is the two-particle phase-space element. 
The hadronic cross section follows by convolution with the parton luminosities,
\begin{align}
d\sigma_{pp \to X_1\ABL}(s) &= \sum_{q = u,c,d,s} \int_{\tau_0}^1 \! d\tau 
	\frac{d\mathcal{L}_{q\bar{q}}^{pp} }{d\tau} 
	d\hat{\sigma}_{\qqbar \to X_i\ABL} (\hat{s}),
\label{eq_hadronicCS_XA}
\end{align}
with the threshold being $\tau_0 = (\mxone + \mabl)^2/s$. 
It is convenient to express the matrix element in terms of the usual Mandelstam invariants,
\begin{align}
\hat{s} = (p_1 + p_2)^2, \quad \hat{t} = (p_1 - p_3)^2, \quad \hat{u} = (p_1 - p_4)^2.
\end{align}
The squared matrix element can then be written as,
\begin{align}
\begin{split}
\left| \overline{\mm}_{\qqbar \to X_1 \ABL}(\hat{s} )\right|^2  &=
	\frac{1}{54} \left( \frac{g_{BL}^2  n_{BL}^X }{2} \right)^2  
	\frac{\hat{t}\hat{u} - \mabl^2\mxone^2}{(\hat{s} - \mzbl^2)^2 + \mzbl^2 \Gamma_{Z_{BL}}^2}
	\cos^2\left( \beta'-\alpha'\right).
\end{split}
\end{align}   

\begin{figure}[t] 
\plot{.5\linewidth}{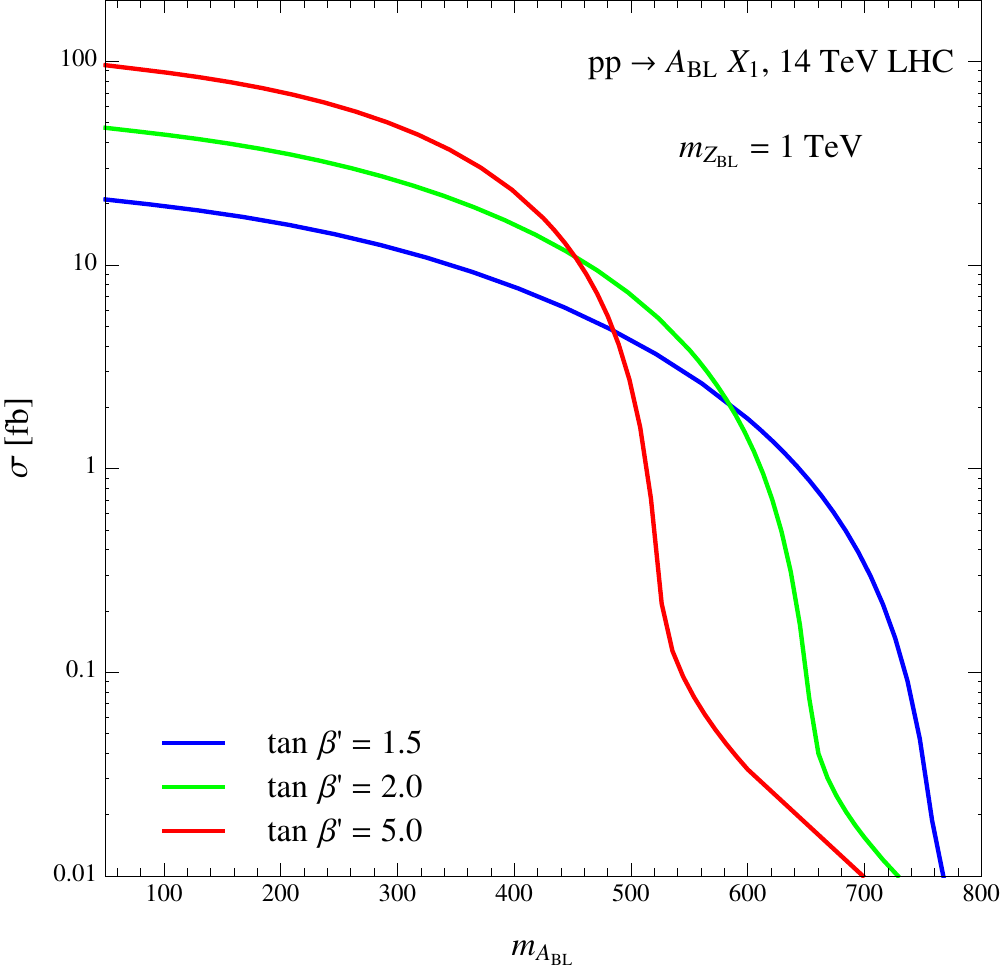}%
\plot{.5\linewidth}{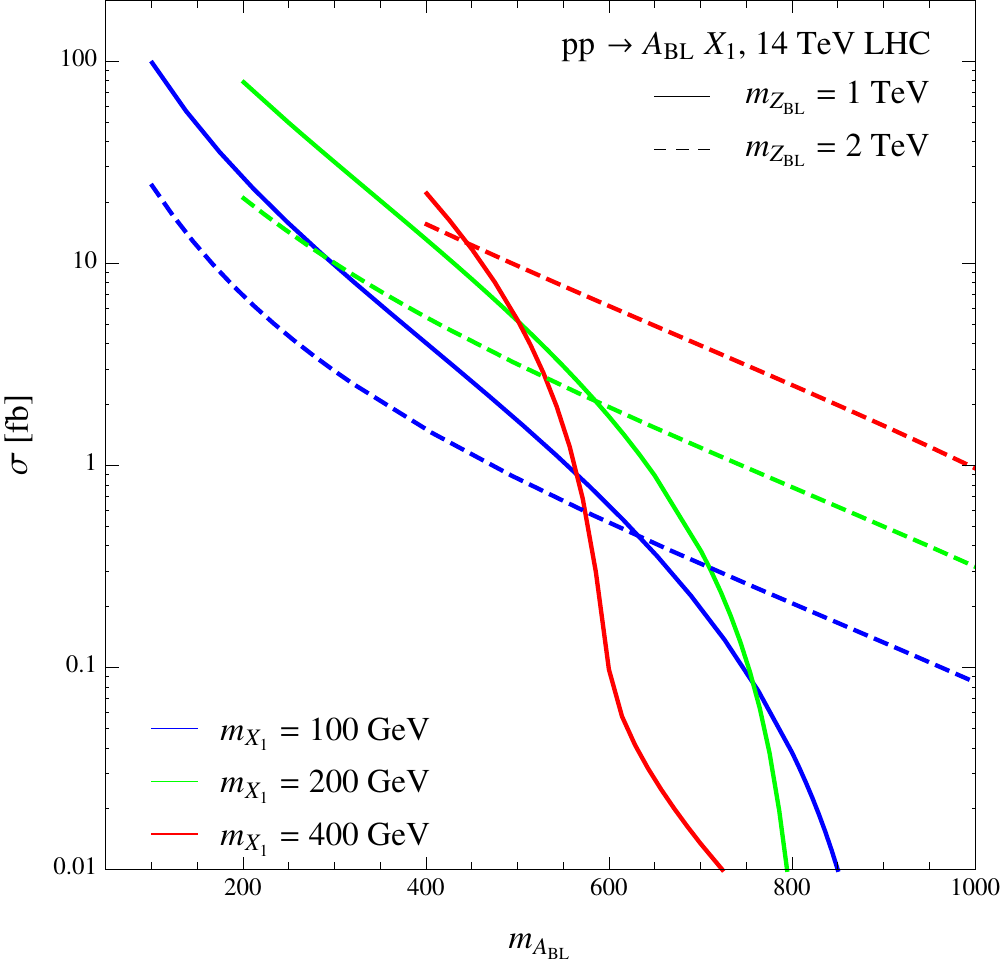}%
\caption{$A_{BL}$ and $X_1$ pair production hadronic cross section for $14$ TeV center of mass energy at the LHC as a function of the mass $\mabl$ for fixed values of $\tan\beta'$ (left) and for fixed values 
of the Higgs mass $\mxone$ (right).  We use the MSTW 2008 LO pdf~\cite{pdf} at a central factorization scale 
$\mu = (\mabl + \mxone)/2$. The suppression due to the threshold $m_{A_{BL}} + m_{X_1} = m_{Z_{BL}}$ is apparent in the solid lines of both plots.  The key feature of this production mechanism is that it is fairly independent of the SUSY spectrum and that it can be large for a sizable part of the parameter space.
}
\label{qqAX.cross.section}
\end{figure}

The numerical cross section results for the pair production of $A_{BL}$ and $X_1$ 
at the LHC at $14$ TeV are shown in Fig.~\ref{qqAX.cross.section} as a function of the mass $\mabl$.  We use the MSTW 2008 LO pdf at a central factorization scale $\mu = (\mabl + \mxone)/2$.  In both plots we use $g_{BL} = m_{Z_{BL}}/(3 \text{TeV})$. In the left panel we plot the pair production cross section versus the mass of $A_{BL}$ for three different values of $\tan \beta'$ for a fixed $Z_{BL}$ mass of 1 TeV.  With these three numbers the entire Higgs sector is fixed and specifically the mass of $X_1$ can be calculated at each point and the larger $\tan \beta'$ the closer $X_1$ and $A_{BL}$ are in mass, with $m_{A_{BL}}>m_{X_1}$.  This specific parameterization has the advantage that the coupling associated with this cross section, $\cos \left(\beta'-\alpha'\right)$ is relatively constant over the range of $m_{A_{BL}}$ shown and so the suppression in the cross section for increased in $m_{A_{BL}}$ is due in small part to the kinematics and in larger part to the $Z_{BL}$ threshold ($m_{A_{BL}} + m_{X_1} = m_{Z_{BL}}$) at which point the $Z_{BL}$ becomes off-shell and the cross section loses its resonance enhancement.

In the right panel, we show the pair production cross section versus $m_{A_{BL}}$ for three different values of $m_{X_1}$ all for two different values of $m_{Z_{BL}}$ ($m_{Z_{BL}} = 1$ TeV solid lines and $m_{Z_{BL}} = 2$ TeV dashed).  The curves start at the point $m_{A_{BL}} = m_{X_1}$ since the CP-even Higgs is at most as heavy as the CP-odd Higgs.  This plot has the advantage of being in terms of the more physical parameter, $m_{X_1}$ but then in this case, the coupling, $\cos \left(\beta' - \alpha'\right)$ changes considerably over the range shown and contributes to the decrease in cross section with increasing $m_{A_{BL}}$, as does the kinematics.

In general the cross section for Higgs pair production can be sizable in a large fraction of the parameter space. For example, when the $\mzbl=1$ TeV, 
$\mabl$ smaller than 500 GeV and $\tan \beta' > 1.5$ the cross section reaches several tens of femtobarn.  Such results are promising for the prospects of testing the mechanism for R-parity conservation.

\begin{figure}[t] 
\plot{.6\linewidth}{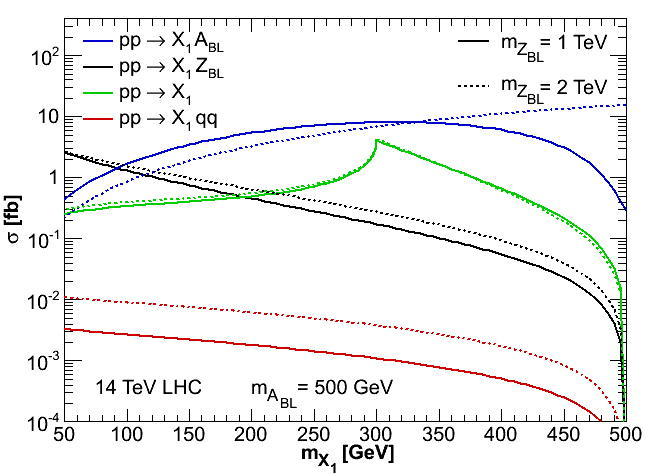}
\caption{Summary of all production channels for $X_1$ production at the LHC 
for  $\mabl = 500$\,GeV (right). One squark is assumed to be light, $m_{\tilde{t}_1} = 150$\, GeV (no mixing),
the heavy neutrino masses are set to $m_{\nu_{R}} = 95$\,GeV. 
We use the MSTW 2008 LO pdf at a central factorization scale (half of the final state masses).}
\label{Allcrosssections}
\end{figure}

In order to complete our analysis we compare in Fig.~\ref{Allcrosssections} all of the possible production mechanisms, including the associate $Z_{BL} X_1$ production and $Z_{BL}$ vector boson fusion. The formulas for the latter two processes are given in Appendix~\ref{app_crosssections}. 
In most of the considered parameter range, the $X_1 \ABL$ pair production dominates and the single 
production (for $m_{\tilde{t}_1} = 150$\,GeV) is at a similar order of magnitude. The cross section for 
associate $Z_{BL} X_1$ production can be large for light particle but drops off quickly for higher masses. The $Z_{BL}$ vector boson fusion, being a 
$2\to3$ particle process, is suppressed from the kinematics and only reaches the $10^{-2}$ fb level.
In terms of testing the mechanism for R-parity conservation the latter two channels play a subleading role and we therefore focus only on the single $X_1$ and $X_1 \ABL$ productions, which have the best potential of shedding light on the stability of the LSP. In the reminder of the paper we will discuss the subsequent decays of the B-L Higgs bosons and the signals for the B-L Higgs production at the LHC.


\section{Higgs Decays and Lepton Number Violating Decays}
The decays of the Higgses depend heavily on the spectrum. Here, we will assume that masses are such that tree-level 
two-body decays are open and dominate. The decays can, of course, be very different in the two models:

{\bf Model I}:

The two-body decays open to $X_1$ are:
\begin{itemize}
	\item $X_1 \to NN$,
	\item $X_1 \to \tilde f \tilde f^*$,
	\item $X_1 \to \tilde {\bar \chi}_i \tilde \chi_j$.
\end{itemize}
Since the coupling of the Higgs to right-handed neutrinos is the defining characteristic of Model I, we will assume for the rest of the paper that only the first channel is open and that the SUSY decays are not kinematically allowed, namely, $m_{X_1} < 2 \, m_{\text{LSP}}$.  For heavier right-handed neutrinos, three general possibilities exist: decay to one right-handed neutrino and one off-shell right-handed neutrino ($m_N < m_{X_1} < 2 \, m_N$), decays into two off-shell right-handed neutrinos ($m_{X_1} < m_N$) or decays similar to Model II. 
Off-shell right-handed neutrinos will manifest as missing energy in final states due to their mixing with the left-handed neutrinos.

The CP-odd scalar, $A_{BL}$, has the following potential two-body decays:
\begin{itemize}
	\item $A_{BL} \to NN$,
	\item $A_{BL} \to \tilde {\bar \chi}_i \tilde \chi_j$.
	\item $A_{BL} \to Z_{BL} X_1$,
\end{itemize}
where the last one would require a heavy $A_{BL}$ outside the reach of the LHC. The two sfermion channel is missing here (compared to $X_1$ decays) since it stems from the $D$-terms in which $A_{BL}$ does not participate. Since $A_{BL}$ is heavier than $X_1$ and we have already assumed that the right-handed neutrino channel is opened to $X_1$ it will also be opened to $A_{BL}$ and we proceed by assuming that all others are closed so that $A_{BL}$ decays $100 \%$ to right-handed neutrinos. 

Therefore, under our assumptions here, the relevant signals to study for Model I are the ones due to the decays of right-handed neutrinos, 
which we will conduct in the next subsection.  As we will see, these decays could lead to lepton number violating signals due to the Majorana nature of the right-handed neutrinos.

{\bf Model II}:

The difference in the second scenario is the lack of the right-handed neutrino--Higgs coupling, thereby only leaving: $S_1 \to \bar{\chi} \chi$ and $S_1 \to \tilde{f}^* \tilde{f}$ as possible tree-level two-body decays. If these are not accessible, one or more of the following decays will dominate:
\begin{itemize}
	\item $S_1 \to \gamma \gamma \quad-\quad$ through a slepton and/or squark loop,
	\item $S_1 \to gg \quad - \quad$ through a squark loop,
	\item $S_1 \to Z_{BL}^* Z_{BL}^* \to$ some combination of leptons and jets (leptons more likely).
\end{itemize}
The CP-odd Higgs now has only two possible decays: $A_{S} \to \tilde{\bar \chi}_i \chi_j$ and $A_{S} \to Z_{BL} S_1$. Both decays are likely to be outside the kinematic range of the light $A_{S}$ we are considering so that one of the final state particles in each of these will have to be off shell.  Since the latter is independent of the SUSY spectrum, we will assume it is the dominate decay with the $Z_{BL}$ off-shell so that the following decays are possible
\begin{itemize}
	\item $A_{S} \to S_1 Z_{BL}^* \to S_1 \ell^\pm \ell^\mp$,
	\item $A_{S} \to S_1 Z_{BL}^* \to S_1 \nu \nu$.
	\item $A_{S} \to S_1 Z_{BL}^* \to S_1 jj$.
\end{itemize}
Therefore, we will assume these three body decays for $A_{S}$ and two body decays for $S_1$,
specifically the scenarios where the lightest Higgs, $S_1$, decays into two sleptons:
\begin{itemize}
	\item $S_1 \to \tilde e \tilde e^*$,
	\item $S_1 \to \tilde \nu^c \tilde \nu^{c*}$.
\end{itemize}
The final states will depend on the identity of the LSP (potentially the dark matter candidate of the universe). We will consider the following possibilities: neutralino, gravitino and right-handed sneutrino and their associated signals.
\subsection{Heavy Neutrinos Decays}
\begin{figure}[h!] 
\plot{.42\linewidth}{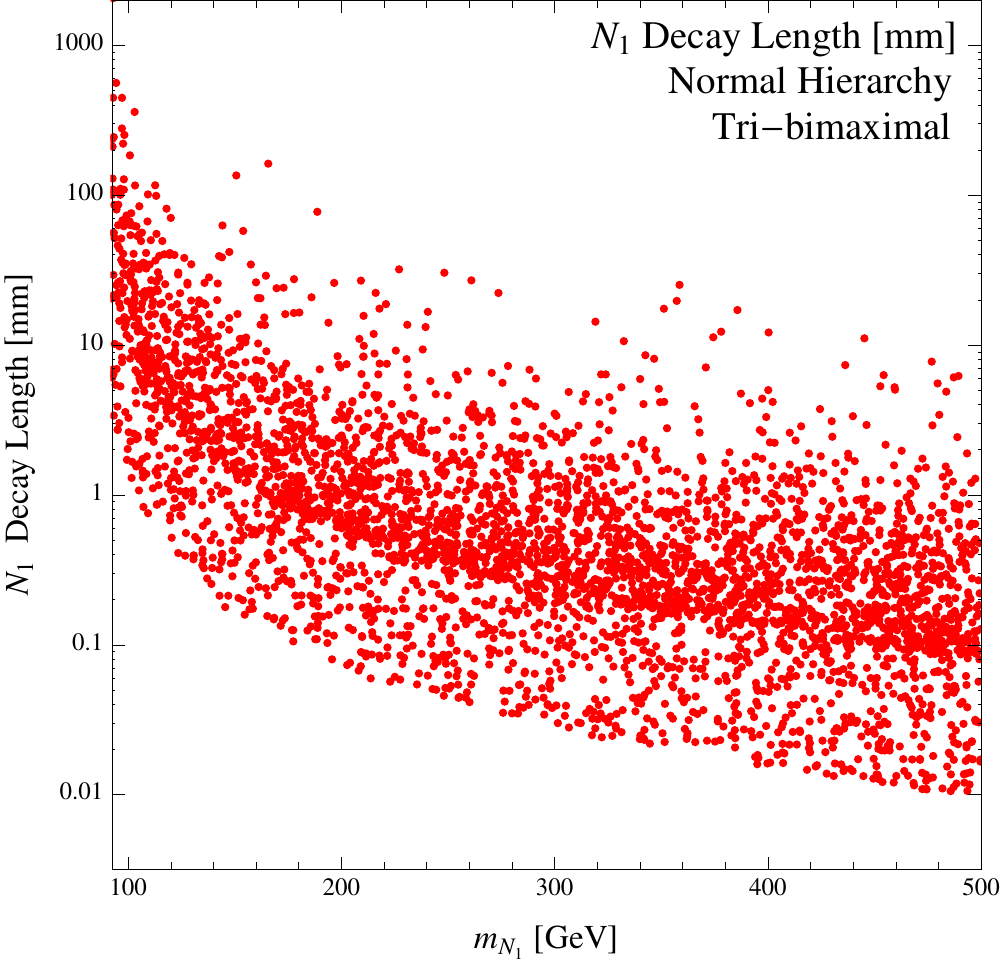}
\plot{.42\linewidth}{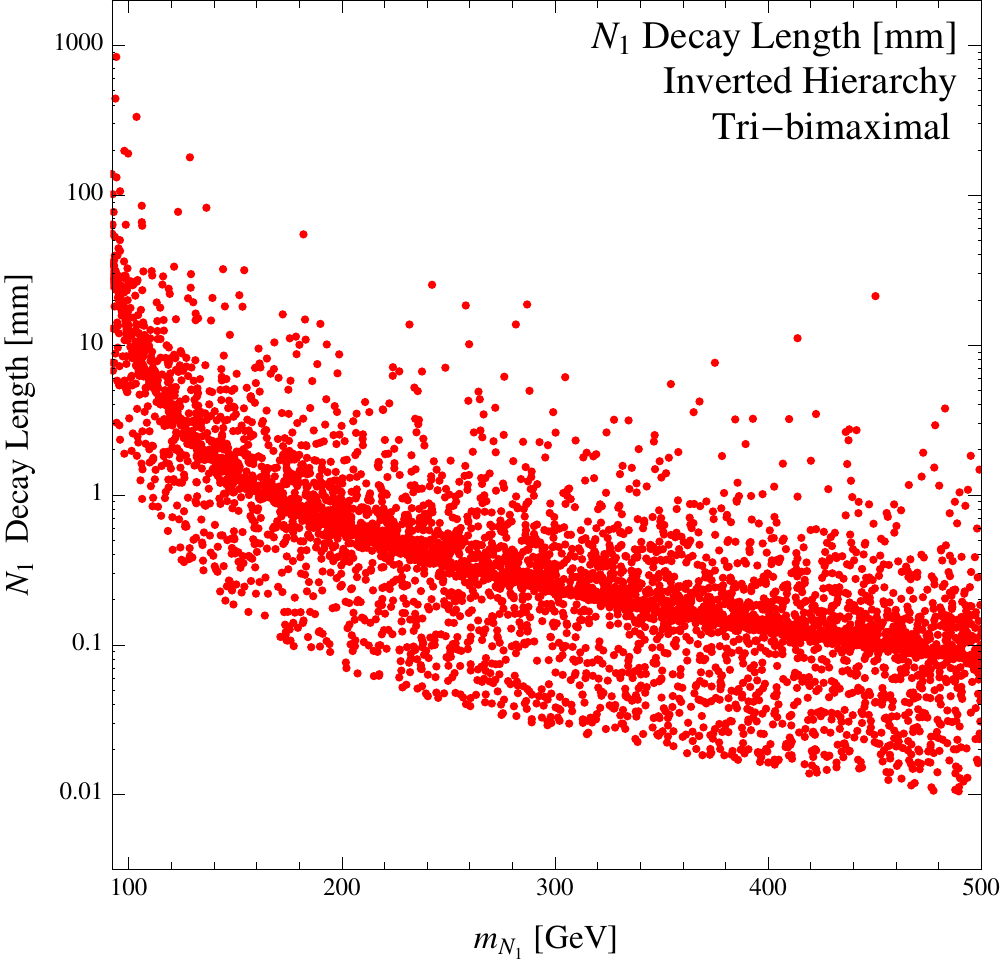}
\plot{.42\linewidth}{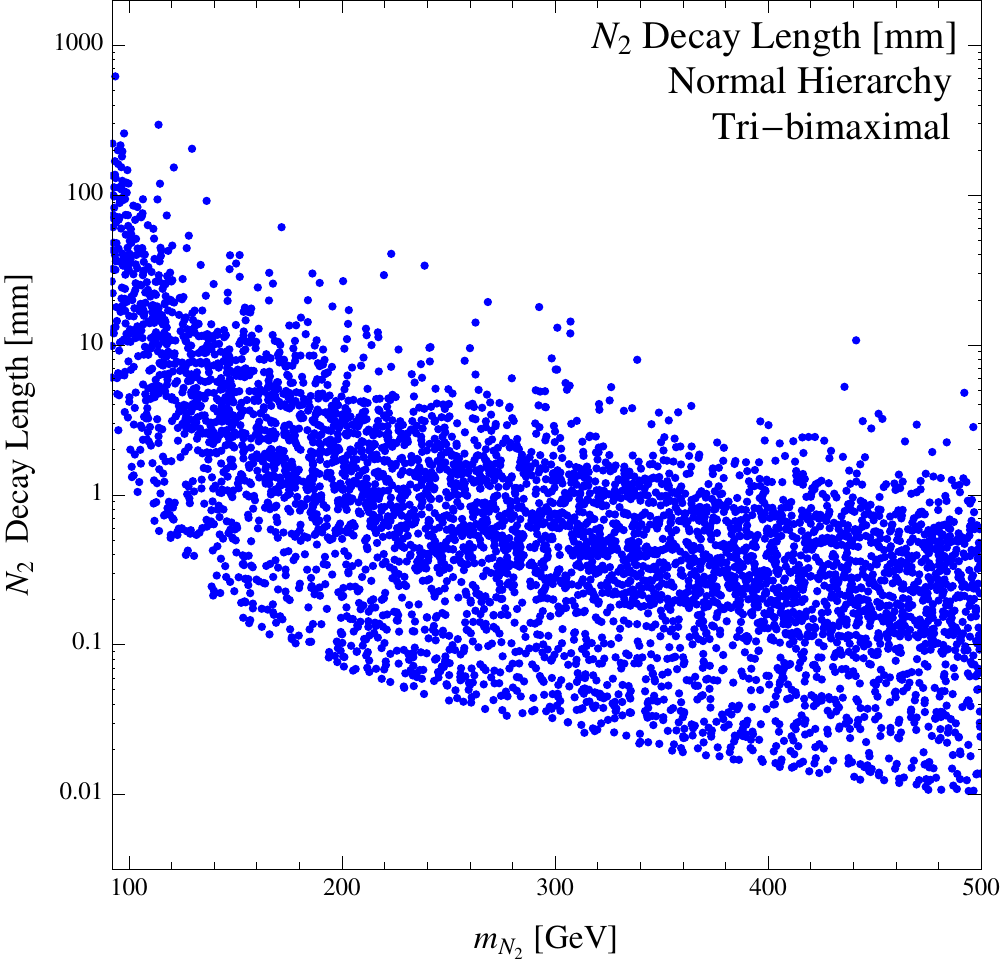}
\plot{.42\linewidth}{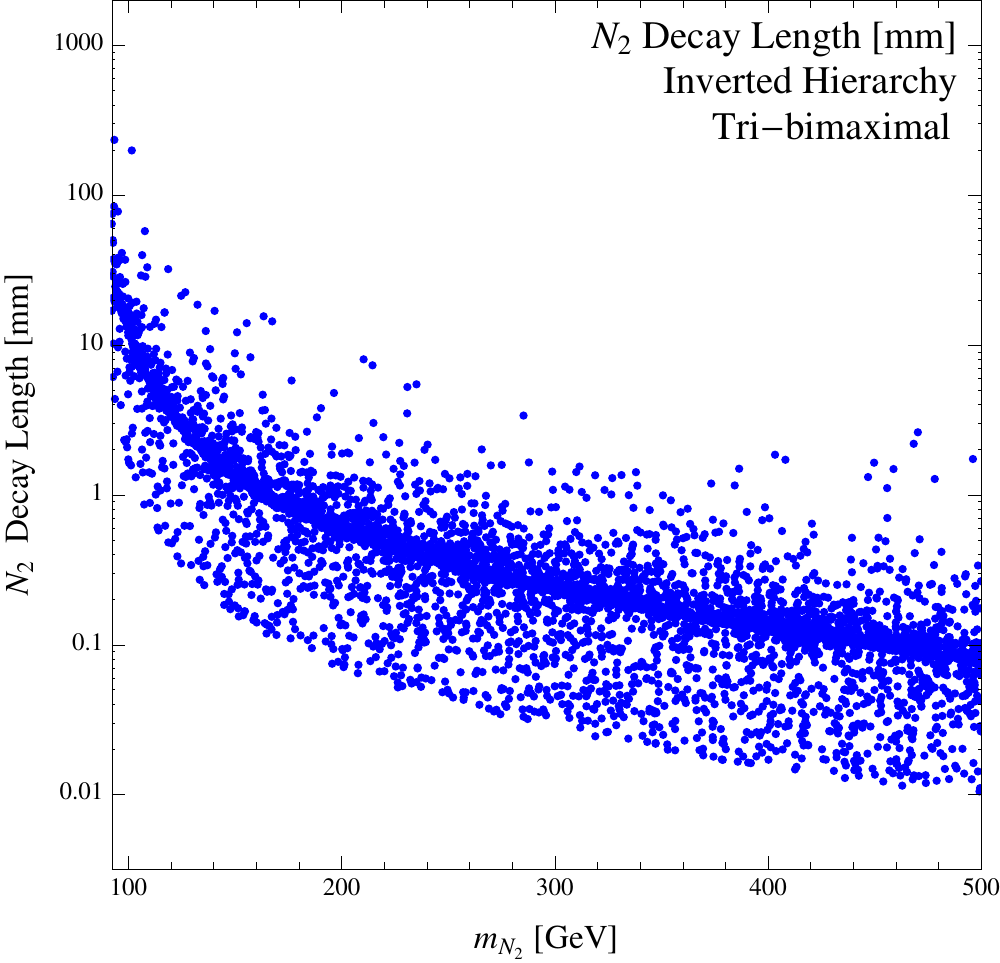}
\plot{.42\linewidth}{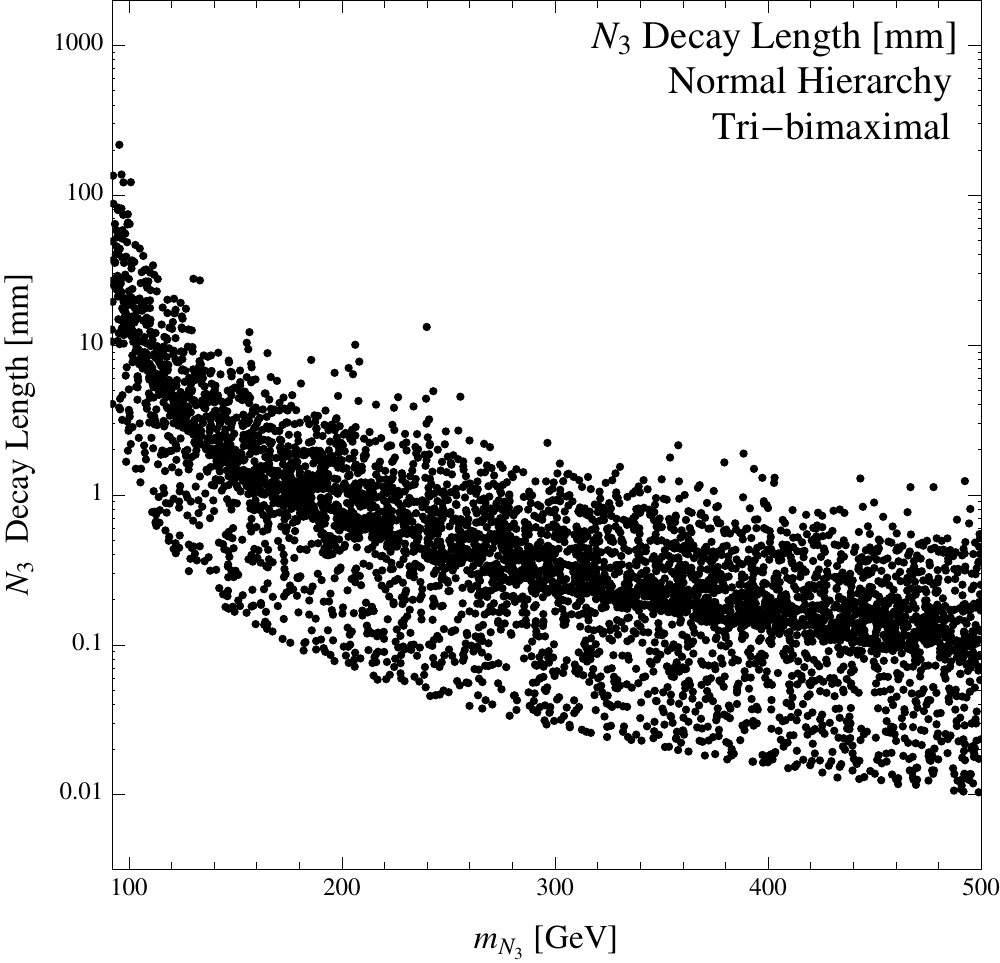}
\plot{.42\linewidth}{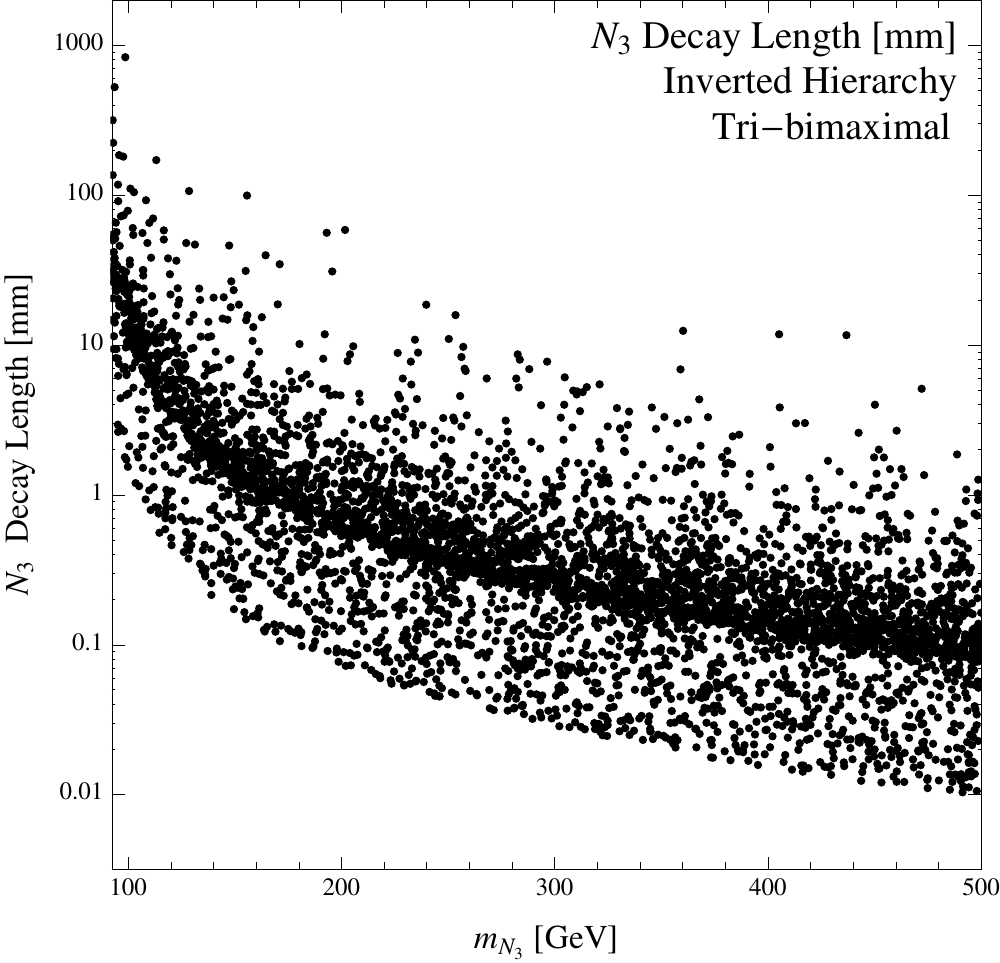}
\caption
{The decay length for the right-handed neutrinos ($N_1$ - red, $N_2$ - blue and $N_3$ - black) versus their mass in the normal hierarchy (inverted hierarchy) on the left (right).  We scan over the following parameters: lightest neutrino mass between $10^{-4}$ eV and $0.4$ eV and $0 \leq \omega_{ij} \leq 1$ for $i,j=1..3$. Due to the small right-handed left-handed neutrino mixing which facilitate these decays, the right-handed neutrinos can be quite long-lived which would lead to displaced vertices.}
\label{Neutrino.Decay}
\end{figure}

Higgses decaying mainly into two right-handed neutrinos allow for
lepton number violating signals due to the subsequent decay of the heavy Majorana right-handed neutrinos.
The leading decay channels for the three heavy neutrinos, $N_a$, include:
\begin{align}
	N_a \to \ell^{\pm}_i W^{\mp},
	\  \
	N_a \to \nu_\ell Z,
	\  \
	N_a  \to  \nu_\ell h_k,
	\ \
	N_a \to \ell^\pm H^\mp.
\end{align}
The amplitude for the two first channels are proportional to the mixing between the leptons and heavy 
neutrinos, while the last one is proportional to the Dirac-like Yukawa terms. While decays to all the MSSM Higgses are possible, typically, only the lightest MSSM Higgs, $h$, is light enough for the scenario we consider here ($m_{N_a} < 500$ GeV) and so we will only take it into account.  The partial decay widths of the heavy Majorana neutrinos $N_i$ are then given by~\cite{TypeI-Zprime,delAguila:2008cj}
\begin{eqnarray}
\Gamma^{\ell W_L}&\equiv&
\Gamma(N_a \to \ell^\pm W_L^\mp)= \frac{g^2}{64\pi
M_W^2}|V_{\ell a}|^2 m_{N_a}^3 \left(1-  \frac{m_W^2}{m_{N_a}^2} \right)^2,
\\
\Gamma^{\ell W_T}&\equiv&\Gamma(N_a \to \ell^\pm W_T^\mp)=
	\frac{g^2}{32\pi  }|V_{\ell a}|^2m_{N_a}    \left(1-  \frac{m_W^2}{m_{N_a}^2} \right)^2,
\\
\Gamma^{\nu_\ell Z_L}&\equiv&\Gamma(N_a\to \nu_\ell Z_L)=\frac{g^2}{
64\pi M_W^2}|V_{\ell a}|^2m_{N_a}^3 \left(1-  \frac{m_Z^2}{m_{N_a}^2} \right)^2,
\\
\Gamma^{\nu_\ell Z_T}&\equiv&\Gamma(N_a \to \nu_\ell Z_T)=\frac{g^2}{
32\pi c_W^2}|V_{\ell a}|^2m_{N_a} \left(1-  \frac{m_Z^2}{m_{N_a}^2} \right)^2,
\\
\Gamma^{\nu_\ell h}&\equiv&\Gamma(N_a \to \nu_\ell h) = 
	\frac{g^2}{64\pi M_W^2}
	|V_{\ell a}|^2
	m_{N_a}^3 \left(1-  \frac{m_h^2}{m_{N_a}^2} \right)^2.
\end{eqnarray}
Here the leptonic mixing between the 
SM charged leptons ($\ell=e, \mu, \tau$) and heavy neutrinos ($N=1,2,3$) reads as~\cite{TypeI-Zprime}:
\begin{eqnarray}
V_{\ell N}= \ V_{PMNS} \ m^{1/2}_\nu \ \Omega \ M^{-1/2}_N,
\label{mixing1}
\end{eqnarray}
a product of matrices related to the neutrino sector.  $V_{PMNS}$ is the PMNS active neutrino mixing matrix. 
Under the assumption that it is real:
\begin{equation}
	V_{PMNS} = 
	\begin{pmatrix}
		c_{12} \, c_{13}
		&
		c_{13} \, s_{12}
		&
		s_{13}
		\\
		-c_{23} \, s_{12}  -   c_{12} \, s_{13} \, s_{23}
		&
		c_{12} \, c_{23} - s_{12} \, s_{13} \, s_{23}
		&
		c_{13} \, s_{23}
		\\
		s_{12} \, s_{23}   -    c_{12} \, c_{23} \, s_{13}
		&
		-c_{12} \, s_{23} - c_{23} \, s_{12} \, s_{13}
		&
		c_{13} \, c_{23}
	\end{pmatrix},
\end{equation}
where $c_{ij} = \cos \theta_{ij}$ and $s_{ij} = \sin \theta_{ij}$ with $0 \leq \theta_{ij} \leq \pi/2$. 
For our numerical predictions we assume the tri-bimaximal ansatz:
\begin{equation}
	s_{12}^2 = \frac{1}{3}, \quad \quad s_{13}^2 = 0, \quad \quad s_{23}^2 = \frac{1}{2}.
\end{equation}
The physical neutrino masses are contained in $m_\nu=diag(m_{\nu_1},m_{\nu_2},m_{\nu_3})$. 
As it is well-known, there are two possible neutrino spectra:
\begin{align}
\begin{split}
	\text{Normal Hierarchy (NH):} & \quad
		m_{\nu_1}, \ \
		m_{\nu_2} = \sqrt{ m_{\nu_1}^2 + \Delta m_{21}^2}, \ \
		m_{\nu_3} = \sqrt{ m_{\nu_1}^2 + |\Delta m_{31}^2| };
	\\
	\text{Inverted Hierarchy (IH):} & \quad
		m_{\nu_1} = \sqrt{m_{\nu_3}^2 + |\Delta m_{31}^2|}, \ \
		m_{\nu_2} = \sqrt{ m_{\nu_1}^2 + \Delta m_{21}^2}, \ \
		m_{\nu_3};
\end{split}
\end{align}
where~\cite{Schwetz:2011qt}
\begin{align}
	7.27 \times 10^{-5} \text{eV}^2 \leq & \, \Delta m_{21}^2 \, \leq 8.03 \times 10^{-5} \text{ eV}^2, 
\\ 
	2.17 \times 10^{-3} \text{ eV}^2 < & \, |\Delta m_{31}^2| \, < 2.54 \times 10^{-3} \text{ eV}^2,
\end{align}
are the solar and atmospheric mass squared differences, respectively. In this paper, we will only use the central value for these masses. 
Finally, $\Omega$~\cite{Casas:2001sr} is a complex orthogonal matrix, which conveniently parameterizes the leftover 
unknown degrees of freedom of the neutrino sector. We shall proceed by 
assuming $\Omega$ to be real. In this case it can be parameterized by three values:
\begin{equation}
	\Omega =
	\begin{pmatrix}
		\sqrt{1 - \omega_{21}^2}
		&
		-\omega_{21}
		&
		0
		\\
		\omega_{21}
		&
		\sqrt{1- \omega_{21}^2}
		&
		0
		\\
		0
		&
		0
		&
		1
	\end{pmatrix}
	\begin{pmatrix}
		\sqrt{1 - \omega_{31}^2}
		&
		0
		&
		-\omega_{31}
		\\
		0
		&
		1
		&
		0
		\\
		\omega_{31}
		&
		0
		&
		\sqrt{1 - \omega_{31}^2}
	\end{pmatrix}
	\begin{pmatrix}
		1
		&
		0
		&
		0
		\\
		0
		&
		\sqrt{1- \omega_{32}^2}
		&
		-\omega_{32}
		\\
		0
		&
		\omega_{32}
		&
		\sqrt{1-\omega_{32}^2}
	\end{pmatrix}.
\end{equation}
Since the penultimate final states of interest are composed of right-handed neutrinos, investigating their decay properties are worthwhile, especially since they depend on the very small neutrino parameter $V_{\ell a}$. In Fig.~\ref{Neutrino.Decay} we do this by plotting the decay length in millimeters versus the mass of the right-handed neutrino ($N_1$ - red, $N_2$ - blue and $N_3$ - black) in the NH (IH) on the left (right). We scan over the unknown parameters: lightest left-handed neutrino mass between $10^{-4}$ eV and $0.4$ eV (where the latter is the upper bound from cosmology) and $0 \leq \omega_{ij} \leq 1$ for $i,j=1..3$. The decay length always increases with decreasing lightest neutrino mass for all other parameters constant.

The noteworthy result from Fig.~\ref{Neutrino.Decay} is that for this range of right-handed neutrino masses--- a mass range chosen to be consistent with a B-L Higgs being produced at the LHC and decay into two on-shell right-handed neutrinos ---the right-handed neutrinos are long-lived (order of millimeter or above) and their decays would exhibit displaced vertices. This is a robust prediction that would lead to spectacular signals and could play a major role in distinguishing these channels. We will expand on this in the next section.

As can be appreciated from the above, the right-handed neutrino decays can be quite different in a given neutrino mass spectrum. To simplify our analysis, we will assume the following: $\Omega = {\mathbf 1}$ and that the right-handed neutrinos are degenerate in mass.  This is in addition to our earlier stated assumptions of tri-bimaximal mixing and central values for the squared mass differences. Ref.~\cite{TypeI-Zprime} studies the effects of varying these values on the decays of the right-handed neutrinos.

Under these assumptions the branching ratios are straightforward and independent of the mass of the lightest neutrino and the neutrino mass hierarchy and are displayed in Table~\ref{N.Br} for degenerate right-handed neutrino masses of $95$ GeV.  These branching ratios would change as the right-handed neutrino mass increases due to the increase in strength of the $Z$ channels and eventually the Higgs channel ---kinematically not allowed for these masses--- and would eventually level off. Clearly, the branching ratios mirror the tri-bimaximal mixing due to $\Omega = {\mathbf 1}$.

\begin{table}[t]
\begin{center}
\begin{tabular}{|c|ccc|}
\hline
 & $N_1$  & $N_2$ & $N_3$
 \\
 \hline
~$\BR(N_i \to e^-\,W^+)$ ~& ~~$31.9 \%$ ~~& ~~$15.9 \%$ ~~&$0 \%$
\\
~$\BR(N_i \to \mu^-\,W^+)$ ~& $8.0 \%$ & $15.9 \%$ & ~~$23.9 \%$~~
\\
~$\BR(N_i \to \tau^-\,W^+)$ ~& $8.0 \%$ & $15.9 \%$ & $23.9 \%$
\\
~$\sum \BR(N_i \to \nu Z)$~ & $4.3 \%$ & $4.3 \%$ & $4.3 \%$
\\
\hline
\end{tabular}
\end{center}
\caption{Branching ratios for the right-handed neutrinos in the special case of tri-bimaximal mixing, central values for the squared mass differences and $\Omega = {\mathbf 1}$ for degenerate right-handed neutrinos masses of $95$ GeV.  }
\label{N.Br}
\end{table}%

These considerations will impact the final states and therefore the signal.  We elaborate on this using the simplifying neutrino sector assumptions mentioned above and focusing on the single $X_1$ production and the pair production of $X_1$ and $A_{BL}$. In both cases, the Higgses decay into two right-handed neutrinos which subsequently decay into two leptons and two heavy vector bosons. The latter will further decay into jets or leptons.  The signals we will focus on in the next section are the one associated with lepton number violation: the two right-handed neutrino decaying into like-sign leptons (muons or electrons) and $W$ bosons, which subsequently decay purely into jets.

\section{Signals at the Large Hadron Collider}
We are now ready to study the possible signals of dynamical R-parity conservation at the LHC. 
For each model, we will outline the final state signals for both the single and pair productions 
and conduct our analysis on the cross section times branching ratio level only.  
We will comment on the relevance of the background but of course our 
comments here would be superseded by a more detailed study of these events.

\underline{\bf Model I}: 

As a reminder, we will proceed under the assumption that both the CP-even and CP-odd Higgses 
decay only into two right-handed neutrinos. This allows for final states consisting of like-sign leptons, 
an indicator of the Majorana nature of the right-handed neutrinos, as long as the $W$ bosons 
decay hadronically. In the case were these decays are not possible, the Higgses could decay 
through off-shell right-handed neutrinos or in a fashion similar to Higgses in Model II 
depending on the spectrum.
 
\underline{Single Production}
\begin{align}
pp \to X_1 \to N \ N  \to e^{\pm}_i W^{\mp} e^{\pm}_j W^{\mp} \to e^{\pm}_i  e^{\pm}_j  4j.
\end{align}
To get a quick naive estimate for the number of events in this channel we do a back of the envelope calculation using
\begin{eqnarray}
N_{2 e 4 j}&\approx& \sigma (pp \to X_1) \times \BR (X_1 \to N_1 N_1 ) \times  2 \BR(N_1 \to e^+ W^- )^2  \times \BR (W \to jj)^2  \times {\cal L}.
\end{eqnarray}
Assuming a large luminosity, ${\cal L}=100 \ fb^{-1}$, and a large cross section of $10 \ fb$ one obtains naively: 
\begin{eqnarray}
N_{2 e 4 j} &\approx& 10 fb \times (1/3) \times  2 (3/10)^2  \times (6/9)^2  \times 100 fb^{-1} =27.
\end{eqnarray}
Then, indicating that a significant number of events can occur.  The exact number of events, $N_{e_n e_m 4j}$, 
can be calculated taking into account the contributions of all the right-handed neutrinos using the following expression:
\begin{align}
\begin{split}
N_{e_n e_m 4j}= \sigma (pp \to X_1) \times & \sum_{a=1\dots3}  
\BR (X_1 \to N_a N_a ) \times  \mathcal{N}^a_{e_n e_m} \times \BR (W \to jj)^2  \times {\cal L},
\end{split}
\end{align}
where $\mathcal{N}^a_{e_n e_m}$ is  the combinatorical factor for two right-handed neutrinos
decaying into $e_n^{\pm} e_m^{\pm}$,
\begin{align}
\mathcal{N}^a_{e_n e_m} &= 2\,  \BR(N_a \to e_n^+ W^- ) \times \BR(N_a \to e_m^+ W^- ) \times \frac{2}{1+\delta_{nm}}.
\end{align} 

We choose a benchmark scenario in order to produce more concrete numbers:
\begin{itemize}
\item Benchmark I:
\begin{itemize}
	\item $m_{A_{BL}} = 1 $TeV, $\ m_{X_1} = 300$ GeV, $\ m_{Z_{BL}} = 1.5$ TeV
	\item $m_{\tilde{t}_1} = 150$ GeV, all other sfermions at $1$ TeV
	\item $m_{N_i} = 95$ GeV for $i=1..3$
	\item In this case $\sigma_{pp \to X_1}  = 16.3$ fb.
\end{itemize}
\end{itemize}
Using these values we display the predicted number of events for the benchmark I in Table~II as well as the combinatorical factor associated with the branching ratios of the right-handed neutrinos to charged lepton final states and $W$ to jets.  This number is independent of the cross section and integrated luminosity and multiplies both to find the number of events.

\begin{table}[t]
\begin{center}
\begin{tabular}{|c||c|c|c|}
\hline
Final State	&	Combinatorics     &  $\ $ Signal$\ $ & $\ $ Background $\ $ 
 \\
\hline
\hline
$2 \, e^\pm \, 4 \, j$	&	0.038	& 62 & 6
\\
$ e^\pm \, \mu^\pm\, 4 \, j$ & 0.030 & 50 & 12
\\
$ \ 2\, \mu^\pm 4 \, j \ $ & 0.027 & 43 & 6
\\
\hline
\hline
\end{tabular}
\caption{\small{Number of events for the three possible two same-sign leptonic final states (with e and $\mu$) for an integrated luminosity of $100 \ \text{fb}^{-1}$ and a single production cross section of $16.3$ fb corresponding to benchmark I with degenerate $95$ GeV right-handed neutrinos.  We also display the combinatoric factor associated with the branching ratio of the Higgs to right-handed neutrinos, right-handed neutrinos to the specific leptonic final state and $W$ bosons to jets.  This factor is independent of the cross section or integrated luminosity and multiplies these two numbers when calculating  number of events.}}
\end{center}
\end{table}%
Meanwhile, the SM background to this sort of signal has been studied before in Ref.~\cite{TypeI-Zprime}, and was found to be dominated by $t \bar{t} W^{\pm}$ production, with a cross section of 4 fb. Using $\BR(t \to j \ell^\pm_i \nu) \sim 10 \%$,  
$\BR(W^\pm \to \ell^\pm_i \nu) \sim 10 \%$ and $\BR(t \to jjj) \sim 67 \%$ we also include an estimate of the number of background events in Table~II.  Two important points are worth noting about the background. The first is that the SM background contains missing energy due to the neutrinos~\cite{TypeI-Zprime}. The second and more important is that, as we saw in the last section, the right-handed neutrinos travel a distance of order millimeters before decaying thereby producing displaced vertices, a further powerful handle on the signal over the background. Therefore, using the information about the two displaced 
vertices in this case one can suppress the SM background. In order to understand the reconstruction of these channels one can use the fact 
that the invariant mass of two jets should be equal to $M_W$, and the invariant mass of two jets and one lepton corresponds to the mass of the right-handed neutrinos~\cite{TypeI-Zprime}.

There is also a possible non-SM background from $Z_{BL} \to NN$, also studied in \cite{TypeI-Zprime}, which would of course have the same signal.  For masses similar to those in benchmark I, the $Z'$ channel will dominate.  Assuming that the $Z_{BL}$ mass is known from the electron or muon channel, the reconstructed mass of the intermediate particle can be used as a handle to differentiate these two channels.  For $Z_{BL}$ masses too heavy for the LHC, the single Higgs production might dominate and act as a complimentary discovery channel for this model.

\underline{Pair Production}

A very important channel is the pair Higgs production through the $B-L$ gauge boson   

\begin{align}
pp \to  Z_{BL}^* \to X_1 A_{BL} \to  N N N N \to  e^{\pm}_i  e^{\pm}_j e^{\pm}_k  e^{\pm}_l   8j,
\end{align}
because it does not depend directly on supersymmetric particles masses and allows for a more reliable signal for this mechanism stabilizing the LSP without depending on the SUSY spectrum.

We again perform a naive estimate to understand the predictions for the number of events with four same sign leptons and eight jets signal using a cross section of $100$ fb and an integrated luminosity of $100 \text{ fb}^{-1}$:
\begin{eqnarray}
N_{4 e 8 j}& \approx& \sigma (pp \to X_1 A_{BL}) \times \BR (X_1 \to N_1 N_1 ) \times  \BR (A_{BL} \to N_1 N_1 ) 
\times 
\nonumber \\ &&
2 \BR(N_1 \to e^+ W^- )^4  \times \BR (W \to jj)^4  \times {\cal L} 
\nonumber \\
 &=& 100 fb \times (1/3) \times (1/3) \times 2 (3/10)^4  \times (6/9)^4  \times 100 fb^{-1} \approx 4.
\end{eqnarray}
Here we pick a second benchmark scenario:
\begin{itemize}
\item Benchmark Scenario II: 
\begin{itemize}
	\item $m_{A_{BL}} = 220$ GeV, $m_{X_1} = 200$ GeV, $m_{Z_{BL}} = 1$ TeV, 
	\item $m_{{\tilde{t}}_1}  =150$ GeV, $m_{\tilde{\tau}_1} = 150$ GeV and all other sfermion at 1 TeV
	\item $\mu_{BL} = 150$ GeV, $M_{BL} = 150$ GeV
	\item $m_{N_i} = 95$ GeV, for $i=1..3$.
	\item The cross is  $\sigma_{pp \to X_1  A_{BL}} =65.7$ fb.
\end{itemize}
\end{itemize}
We display the predicted number of events in Table~III for the five possible final states with e and $\mu$ leptons. 
We also show the combinatorics factor which takes into account the branching ratios of the Higgses into 
right-handed neutrinos, right-handed neutrinos into leptons and $W$ bosons to jets. This number can be multiplied 
by the cross section and integrated luminosity to yield the number of events. Note that this second benchmark 
is in some ways complimentary to benchmark I with respect to the mass of $A_{BL}$.

\begin{table}[t]
\begin{center}
\begin{tabular}{|c||c|c|}
\hline
Final State	& Combinatorics     &  $\ $ Number of Events $\ $
 \\
\hline
\hline
$4 \, e^\pm \, 8 \, j$	&	0.00072	& 4.8
\\
$ 3\, e^\pm \, \mu^\pm\, 8 \, j$ & 0.0012 & 7.6
\\
$ \ 2\, e^\pm \, 2 \mu^\pm\, 8 \, j \ $ & 0.0015 & 9.7
\\
$ \, e^\pm \, 3 \mu^\pm\, 8 \, j$ & 0.00081 & 5.3
\\
$4 \mu^\pm\, 8 \, j$ & 0.00035 & 2.3
\\
\hline
\end{tabular}
\caption{\small{Number of events for the five possible four same-sign leptonic final states for an integrated luminosity of $100 \ \text{fb}^{-1}$ and a pair production cross section of $65.7$ fb corresponding to benchmark II (degenerate $95$ GeV right-handed neutrinos).  We also display the combinatorics factor which combines the branching ratios for the Higgses into right-handed neutrinos, right-handed neutrinos to specific leptonic final states and $W$ bosons into jets.  This factor is independent of the cross section or integrated luminosity and simply multiplies any cross section and integrated luminosity to give the total number of events.}}
\end{center}
\label{NE}
\end{table}%

In this case the main SM background is $t \bar{t} W^{\pm} t \bar{t} W^{\pm}$ which has a negligible cross section. 
It is important to mention that in this case one has four displaced vertices making the signal quite special.
This does not change the fact though that the reconstruction in this case is quite challenging due to the presence 
of eight jets in the final state. Imposing the condition that the invariant mass of two jets, $|M(jj)-M_W | < 15 $ GeV~\cite{TypeI-Zprime}, 
can improve the reconstruction process as well as the order millimeter displaced vertices due the long lifetimes of the right-handed neutrino.
A more detailed study will be considered in a future publication.

\underline{\bf Model II}:

In this section we assume that $n_{\phi}=4$ and that the CP-even Higgs $S_1$, decays into two sfermions while the CP-odd Higgs, $A_S$ decays into $S_1$ and opposite-sign lepton pairs from an off-shell $Z_{BL}^*$ (jets and neutrinos are also possible).  Regardless of these concrete assumption, the signals still depends on the SUSY spectrum.  We will therefore only highlight some interesting scenarios and finish by addressing some alternatives to the two-body sfermion decays. 

\begin{itemize}

\item $S_1 \to \tilde e \tilde e^*$: $\ \tilde{\chi}_1$ as the LSP:

Here we assume that $m_{\tilde{\chi}_1} \ < \ m_{\tilde{e}} \ < \ m_{S_1}/2 $ and that the lightest neutralino is the LSP (not necessarily 
a B-L neutralino) and that $\tilde{e}$ is the NLSP.

\underline{Single Production}:
In the case of the single production one has
\begin{align}
pp \to S_1 \to \tilde{e}^* \  \tilde{e}  \to  e^{\pm} e^{\mp} \tilde{\chi}_1 \tilde{\chi}_1,
\end{align}
and it yields opposite-sign lepton and missing energy. For benchmark I, where the single production cross section is independent of $n_\phi$ so that the single production cross section is unchanged ($\sigma_{pp \to S_1} =16.3$ fb), and the number of events assuming $100 \%$ branching ratios and ${\cal L}=1 \  \textrm{fb}^{-1}$ is
\begin{align}
N_{e^{\pm} e^{\mp} E_T^{miss}} = 
\sigma_{pp \to S_1} \times \BR(S_1 \to \tilde{e}^* \  \tilde{e}) \times \BR(\tilde{e} \to  e \tilde{\chi}_1 )^2 \times {\cal L} \sim 16.
\end{align}
Notice that this estimation is naive because one has to assume a large branching ratio for the decays into selectrons.
The main SM background is the $WW, ZZ$, and  $t \bar{t}$ production, but cutting on large missing $E_T^{miss}$ 
one could reduce this background. See Refs.~\cite{Tao-DM,DM-review} for recent studies and examples of different techniques.

\underline{Pair Production}: As we discussed above the pair production can give us a better idea of the cross section without assuming a particular supersymmetric spectrum. In this case one can have the following signals:
\begin{align}
pp \to S_1 A_S \to \tilde{e}^* \  \tilde{e} \  S_1 e^+_i e^-_i  \to  e^{\pm} e^{\mp} e^{\pm} e^{\mp} e^+_i e^-_i \tilde{\chi}_1 \tilde{\chi}_1 \tilde{\chi}_1 \tilde{\chi}_1.
\end{align}
Then, in this case one has three pairs of leptons and missing $E_T^{miss}$. This cross section does depend on the value 
of $n_\phi$, and for benchmark II, $\sigma_{pp \to S_1 A_S} = 160$ fb when $n_\phi = 4$ (note that the $Z_{BL}$ width also changes with $n_\phi$ so that the cross section doesn't simply scale with this parameter). 
The number of events for ${\cal L}=1 \  \textrm{fb}^{-1}$ can be estimated as
\begin{align}
	N_{3 (e^{\pm} e^{\mp}) E_T^{miss}} = \sigma_{pp \to S_1 A_{S}} \times \BR (Z_{BL} \to e^+_i e^-_i) \times {\cal L} \sim 40,
\end{align}
where $\BR (Z_{BL} \to e^+_i e^-_i) \sim 25 \%$ in this case.
As in the single production case, demanding a large missing $E_T$ one should be able to reduce the background 
which is much less severe since it involves more gauge fields or three pairs of top quarks.

\item Gravitino LSP and $\tilde{\chi}_1$ NLSP:

For which we assume the hierarchy: $m_{\tilde{G}} < m_{\tilde{\chi}_1} \ < \ m_{\tilde{e}} \ < \ m_{S_1}/2 $ 
and that neutralino decays within the detector: $\tilde{\chi}_1 \to \gamma \tilde G$.

\underline{Single Production}:
\begin{align}
pp \to S_1 \to \tilde{e}^* \  \tilde{e}  \to  e^{\pm} e^{\mp} \gamma \gamma \tilde{G} \tilde{G},
\end{align}
is marked by a pair of opposite-sign leptons, photons and missing $E_T$. The number of events is the 
same as in the previous scenario since we assuming that all branching ratios are $100 \%$. 
For benchmark I and $\mathcal{L} = 1 \text{ fb}^{-1}$:
\begin{align}
N_{e^{\pm} e^{\mp} \gamma \gamma E_T^{miss}} = \sigma_{pp \to S_1} \times {\cal L} \sim 16.
\end{align}
Again, this is a naive estimation of the number of events. It is easy to see that one can satisfy the new bounds from CMS 
on gauge mediation~\cite{CMSboundGM}.
  
\underline{Pair Production}:
One can have channels with multileptons and multiphotons if one uses the Higgs pair production:
\begin{align}
	pp \to S_1 A_S \to \tilde{e}^* \  \tilde{e} \  S_1 e^+_i e^-_i  \to 
		e^{\pm} e^{\mp} e^{\pm} e^{\mp} e^+_i e^-_i \gamma \gamma \gamma \gamma
		\tilde G \tilde G \tilde G \tilde G
\end{align}
produces multileptons and multiphotons and missing $E_T$. The number of events is given by
\begin{align}
N_{3 (e^{\pm} e^{\mp}) 4 \gamma E_T^{miss}} = \sigma_{pp \to S_1 A_{BL}} \times \BR (Z_{BL} \to e^+_i e^-_i) \times {\cal L} \sim 40,
\end{align}
for benchmark II, $n_\phi = 4$ and $1 \text{ fb}^{-1}$ of integrated luminosity. One can see that these results are 
in agreement with the bounds from CMS~\cite{CMSboundGM}.

\item $\tilde{\nu}^c$ as the LSP:

In this case the right-handed sneutrino can be a dark matter candidate and in principle 
the Higgs $S_1$ can decay mainly into dark matter, while $A_{S}$ decays into two leptons and dark matter, 
$A_S \to S_1 Z_{BL}^* \to S_1 e^{+}_i e^-_i \to  (\tilde{\nu}^c)^* \tilde{\nu}^c e^{+}_i e^-_i $. 
The number of events can be estimated as in the previous cases. However, since in order to study these channels one needs 
to make use of the initial state radiation (ISR), we postpone this study for a future publication.
For a study on sneutrino dark matter in this context see Ref.~\cite{Khalil:2008ps}.

\item LSP heavier than $S_1$:

Here again there is a lot of variability depending on the specific spectrum. 
We simply refer the reader to Table~\ref{other.decays} for the possible final states 
in this case and leave the calculations for the number of events to a future paper.
\begin{table}[t]
\caption{Channels with multileptons and multiphotons in Model II when $m_{LSP} > m_{S_1}$.}
\begin{center}
\begin{tabular}{|c||c|c|}
\hline
$S_1$ Decay & Single production final state & Pair production final state
\\
\hline
\hline
$S_1 \to \gamma \gamma$ & $\gamma \gamma$ & $\ell^\pm_i \ell^\mp_i \gamma \gamma \gamma \gamma$
\\
$S_1 \to jj$ & $jj$ & $\ell_i^\pm \ell_i^\mp jjjj$ 
\\
$S_1 \to \ell^\pm_j \ell^\mp_j \ell^\pm_k \ell^\mp_k$ & $\ell^\pm_j \ell^\mp_j \ell^\pm_k \ell^\mp_k$ &
$\ell^\pm_i \ell^\mp_i \ell^\pm_j \ell^\mp_j \ell^\pm_k \ell^\mp_k \ell^\pm_l \ell^\mp_l \ell^\pm_m \ell^\mp_m$
\\
\hline
\end{tabular}
\end{center}
\label{other.decays}
\end{table}%
It is important to mention again that in the case of Model II it is not possible to make well-defined predictions for the signals 
because we do not know the SUSY spectrum. If one sticks to a particular SUSY breaking scenario one could see which of these signals 
are the relevant ones. In this paper we pointed out the different possibilities and a detailed study is beyond the scope of this article. 
\end{itemize}


\section{Summary}
The possibility to test the mechanism responsible for the stability of the lightest 
supersymmetric particle at the Large Hadron Collider has been investigated in detail. 
As it has been discussed in this article, the simplest theoretical frameworks 
where R-parity conservation can be explained dynamically are based 
on B-L gauge symmetry. We discuss two different models and find 
the following interesting results:

\begin{itemize}

\item In the simplest theoretical frameworks where one can explain dynamically the conservation of R-parity 
one must have new Higgs bosons which decay mainly into two right-handed neutrinos or into two sfermions.

\item We have investigated the production mechanisms and decays of the B-L Higgses. 
We have found that the Higgs pair production mechanism is quite relevant 
for the testability of the mechanism for R-parity conservation, because its predictions 
are independent of the supersymmetric spectrum.

\item In Model I, where the B-L Higgs couples at tree level to the right-handed neutrinos, one can have lepton 
number violating signals with multileptons and multijets. In this case, if the masses of  the new Higgses are 
below $500$ GeV one obtains multiple displaced vertices due to the presence of long-lived right-handed neutrinos.  

\item A new class of models for the dynamical conservation of R-parity has been discussed. In this case the new 
physical Higgses couple only to the sfermions at tree level and the neutrinos are Dirac fermions. One finds 
different exotic signals. However, those channels depend on the supersymmetric spectrum. 
In a simple scenario, such as gauge mediation, one can have channels with multileptons and multiphotons.

\end{itemize}

The testability of the mechanism for R-parity conservations may help us to understand the link between missing energy at the LHC 
and the cold dark matter of the universe.
  
\subsection*{Acknowledgment}
This work was supported in part by the US DOE under contract No. DE-FG02-95ER40896.
\appendix
\section{Feynman Rules and Cross Sections}

\subsection{Feynman Rules}
\begin{itemize} 

\item \underline{Higgs-squark-squark}:
\begin{align}
	X_1 \tilde q_1^* \tilde q_1: & \quad 
	i g^{X_1}_{\squark_1~\squark_1} = 
	\frac{i}{6} \, g_{BL} \, m_{Z_{BL}} \  \sin (\alpha' + \beta') \  \cos 2 \theta_{\tilde q},
	\\
	X_1 \tilde q_2^* \tilde q_2: & \quad 
	i g^{X_1}_{\squark_2~\squark_2} = 
	-\frac{i}{6} \, g_{BL} \, m_{Z_{BL}} \  \sin (\alpha' + \beta')  \  \cos 2 \theta_{\tilde q},
	\\
	X_1 \tilde q_1^* \tilde q_2: & \quad 
	i g^{X_1}_{\squark_1~\squark_2} = 
	-\frac{i}{6} \, g_{BL} \, m_{Z_{BL}} \  \sin (\alpha' + \beta' ) \  \sin 2 \theta_{\tilde q},
	\\
	X_2 \tilde q_1^* \tilde q_1: & \quad 
	i g^{X_2}_{\squark_1~\squark_1} = 
	-\frac{i}{6}\, g_{BL} \, m_{Z_{BL}} \  \cos (\alpha' + \beta') \  \cos 2 \theta_{\tilde q},
	\\
	X_2 \tilde q_2^* \tilde q_2: & \quad 
	i g^{X_2}_{\squark_2~\squark_2} = 
	\frac{i}{6}\, g_{BL} \, m_{Z_{BL}} \   \cos (\alpha' + \beta') \  \cos 2 \theta_{\tilde q},
	\\
	X_2 \tilde q_1^* \tilde q_2: & \quad 
	i g^{X_2}_{\squark_1~\squark_2} = 
	\frac{i}{6}\, g_{BL} \, m_{Z_{BL}}  \   \cos (\alpha' + \beta')  \  \sin 2 \theta_{\tilde q},
\end{align}
for any squark $\tilde q$.

\item \underline{Higgs-fermion-fermion}:
\begin{align}
	X_1 N_i N_i: & \quad 
	i \, 2 \sqrt 2 \, f_i \cos \alpha'= -2 \, i \, g_{BL} \frac{\cos \alpha'}{\sin \beta'} \frac{m_{N_i}}{m_{Z_{BL}}},
	\\
	X_2 N_i N_i: & \quad 
	-i \, 2 \sqrt 2 \, f_i \sin \alpha' = 2 \, i \, g_{BL} \frac{\sin \alpha'}{\sin \beta'} \frac{m_{N_i}}{m_{Z_{BL}}},
	\\
	A_{BL} N_i N_i: & \quad
	2 \sqrt 2 \, f_i \cos \beta' \gamma_5 = -2 \, g_{BL}\frac{1}{\tan \beta'} \frac{m_{N_i}}{m_{Z_{BL}}},
\end{align}

\item \underline{quark-squark-gluino}:
\begin{align}
	g^{a\mu} \tilde q^*_\alpha(p) \tilde q_\beta(k): & \quad
		-i g_3(p+k)^\mu \lambda^a_{\alpha \beta},
	\\
	g^{a \mu} g^{b \nu} \tilde q^*_\alpha q_\beta: & \quad
		i g_3^2 \left(\lambda^a \lambda^b + \lambda^b \lambda^a\right)_{\alpha \beta} g^{\mu \nu},
	\\
	g^{c \mu} \tilde g^{b\dagger} \tilde g^a: & \quad
		-g_3 f_{abc} \gamma^\mu,
	\\
	q_\alpha \tilde q^*_{1\beta} \tilde g^a: & \quad
		i g_3 \sqrt{2} \lambda^a_{\alpha \beta} \left(\cos \theta_{\tilde q} P_L +  \sin \theta_{\tilde q} P_R\right),
	\\
	q_\alpha \tilde q^*_{2\beta} \tilde g^a: & \quad
		i g_3 \sqrt{2} \lambda^a_{\alpha \beta} \left(- \sin \theta_{\tilde q} P_L + \cos \theta_{\tilde q} P_R\right),
\end{align}
where $\lambda^a$ are the generators of $SU(3)$, $\alpha \text{ and } \beta$ represent color and $f_{abc}$ are the structure constants for $SU(3)$.

\item $Z_{BL} Z_{BL} \phi_i$:
\begin{align}
	Z_{BL}^\mu Z_{BL}^\nu \phi_1: & \quad -i g_{BL} n_{BL} m_{Z_{BL}} \sin (\beta'-\alpha') g_{\mu \nu},
	\\
	Z_{BL}^\mu Z_{BL}^\nu \phi_2: & \quad -i g_{BL} n_{BL} m_{Z_{BL}} \cos (\beta'-\alpha') g_{\mu \nu},
\end{align}

\item $Z_{BL} \phi_i  A_{BL}$:
\begin{align}
	Z_{BL}^\mu \phi_1 A_{BL}: & \quad g_{BL} \frac{n_{BL}}{2} \cos (\alpha'-\beta') (p_1-p_2)_\mu,
	\\
	Z_{BL}^\mu \phi_2 A_{BL}: & \quad g_{BL} \frac{n_{BL}}{2} \sin (\alpha'-\beta') (p_1-p_2)_\mu,
\end{align}

\item $Z_{BL} \bar{f} f$: Here $f=u,d,e$.
\begin{align}
	Z_{BL}^\mu \bar{f} f: & \quad -i g_{BL} \frac{n_{BL}^f}{2} \gamma_\mu,
\end{align}
\item $Z_{BL} \bar{\nu} \nu$ and $Z_{BL} \bar{N} N$: Here $N=\nu_R + (\nu_R)^C$ , 
\begin{align}
	Z_{BL}^\mu \bar{\nu} \nu: & \quad i \frac{g_{BL}}{2} \gamma_\mu \gamma_5,
\end{align}
\begin{align}
	Z_{BL}^\mu \bar{N} N: & \quad i \frac{g_{BL}}{2} \gamma_\mu \gamma_5,
\end{align}
\item $Z_{BL} \tilde{f}^\dagger_i \tilde{f}_j$: Here $\tilde{f}_i= \tilde{u}_i^a, \tilde{d}_i^a, \tilde{e}_i^a, \tilde{\nu}_i^a$, where $i,j=1,2$ and $a=1,2,3$.
\begin{align}
	Z_{BL}^\mu \tilde{f}^\dagger_i \tilde{f}_j: & \quad - i \frac{g_{BL}}{2} n_{BL}^f \left( p_1 -p_2\right)_\mu \, \left( U_{i1}^{\tilde{f}} U_{j1}^{\tilde{f}} + U_{i2}^{\tilde{f}} U_{j2}^{\tilde{f}} \right)^2.
\end{align}
\item $Z_{BL} \bar \chi_i \chi_j$
\begin{align}
	Z_{BL}^\mu \bar \chi_i \chi_j: & \quad
	-i g_{BL} \frac{\eta_{BL}}{2}
	\left(
		N_{i \tilde{\bar X}} N^\dagger_{\tilde{\bar X} j} - N_{i \tilde X} N^\dagger_{\tilde Xj}
	\right)
	\gamma_\mu \frac{\gamma_5}{2} \left(1+ \delta_{ij}\right).
\end{align}
\end{itemize}
%
\subsection{Cross Sections} 
\label{app_crosssections}
The process $pp \to Z_{BL}^* \to X_1  Z_{BL}$ is described as  
associated production or Higgs strahlung. 
The differential partonic cross sections is given by
\begin{align}
d\hat{\sigma}_{\qqbar \to X_1 \ZBL} (\hat{s}) & =
	 \left| \overline{\mm}_{\qqbar \to X_1\ZBL }(\hat{s} )\right|^2 
	 \frac{d\text{PS}^{(2)}}{2\hat{s}},
\end{align}
in terms of the matrix element,
\begin{align}
\begin{split}
\left| \overline{\mm}_{\qqbar \to X_1 \ZBL}(\hat{s})\right|^2  &=
	\frac{1}{54} \left( \frac{g_{BL}^2  n_{BL}^X }{2} \right)^2  
	\frac{\mzbl^2 \hat{s} + (\hat{t} - \mzbl^2) (\hat{u} - \mzbl^2)}{(\hat{s} - \mzbl^2)^2 + \mzbl^2 \Gamma_{Z_{BL}}^2}
	\sin^2\left( \beta'-\alpha'\right).
\end{split}
\end{align}   
The hadronic cross section follows by convolution in analogy to Eq.\eqref{eq_hadronicCS_XA}
with the production threshold being $\tau_0 = (\mxone + \mzbl)^2/s$.

The result for the $\ZBL$ boson fusion, $q(p_1) q'(p_2) \to q(p_3) X_1 (p_4) q'(p_5)$,
 arises from the diagram shown in Fig.~\ref{fig_production}(d) and the one 
with crossed external quark lines. In terms of extended Mandelstams, 
$\hat{t}_{1i} = (p_1 - p_i)^2$ and $\hat{u}_{2i} = (p_2 - p_i)^2$, we can write for the squared matrix element,
\begin{align}
\begin{split}
& \left| \overline{\mm}_{qq' \to  X_1 qq'} (\hat{s} )\right|^2  =
	\frac{2}{9} \left( \frac{g_{BL}^3  n_{BL}^X }{2} \right)^2  \mzbl^2 \sin^2\left( \beta'-\alpha'\right) 
\\
\times\Bigg\lbrace 
& \left[\hat{s}^2 + \hat{s}(\hat{t}_{14} + \hat{u}_{24} - \hat{u}_{23}) 
	- (\hat{t}_{13}+\hat{t}_{14})\hat{u}_{23}
	+ m_{X_1}^2 (\hat{u}_{23} -\hat{s})\right] 
	\frac{1}{(\hat{u}_{25}-\mzbl^2)^2 (\hat{t}_{13} -\mzbl^2)^2}
\\
+ & \left[\hat{s}^2 + \hat{s}(\hat{u}_{24} + \hat{t}_{14} - \hat{t}_{13}) 
	- (\hat{u}_{23}+\hat{u}_{24})\hat{t}_{13}
	+ m_{X_1}^2 (\hat{t}_{13} -\hat{s})\right] 
	\frac{1}{(\hat{t}_{15}-\mzbl^2)^2 (\hat{u}_{23} -\mzbl^2)^2}
\\
+ & \left[\frac{3}{2}\hat{s}( \hat{s} + \hat{t}_{14} + \hat{u}_{24} - m_{X_1}^2 )\right] 
	\frac{1}{(\hat{u}_{25}-\mzbl^2)(\hat{t}_{15}-\mzbl^2) (\hat{t}_{13} -\mzbl^2)(\hat{u}_{23} -\mzbl^2)}.
\Bigg\rbrace
\end{split}
\end{align} 
The matrix element does not depend on the electric charge or the flavor of the quarks and at the hadronic level we can just sum over 
all possible initial states by adding the respective parton densities:
\begin{align}
d\sigma_{PP \to X_i qq'}(s) &= \sum_{q,q' = u,c,d,s} \int_{\tau_0}^1 \! d\tau 
	\left( \frac{d\mathcal{L}_{qq'}^{PP} }{d\tau} +\frac{d\mathcal{L}_{q\bar{q}'}^{PP}}{d\tau} + \frac{d\mathcal{L}_{\bar{q}\bar{q}'}^{PP}}{d\tau} \right)
	\, \left| \overline{\mm}_{qq' \to  X_1 qq'} (\hat{s} )\right|^2 \,
	\frac{1}{2\hat s} \, d\text{PS}^{(3)},
\end{align}
where $d\text{PS}^{(3)}$ is the 3-particle phase-space element 
and the production threshold is $\tau_0 = \mxone^2/s$.   




\begin{thebibliography}{99}

\bibitem{SUSY}
M. Drees, R. Godbole and P. Roy, ``Theory and Phenomenology
of Sparticles," (World Scientific, 2004).

\bibitem{Aulakh:1982yn}
  C.~S.~Aulakh and R.~N.~Mohapatra,
  ``Neutrino As The Supersymmetric Partner Of The Majoron,"
  Phys.\ Lett.\  B {\bf 119}, 136 (1982).

\bibitem{Hayashi:1984rd}
  M.~J.~Hayashi and A.~Murayama,
  ``Radiative Breaking of $SU(2)_R \times U(1)_{B-L}$ gauge symmetry induced by broken
  N=1 Supergravity in a Left-Right symmetric model,''
  Phys.\ Lett.\  B {\bf 153}, 251 (1985).

\bibitem{Krauss:1988zc}
  L.~M.~Krauss, F.~Wilczek,
  ``Discrete Gauge Symmetry in Continuum Theories,'
  Phys.\ Rev.\ Lett.\  {\bf 62}, 1221 (1989);
  S.~P.~Martin,
  ``Some simple criteria for gauged R-parity,''
  Phys.\ Rev.\  D {\bf 46}, 2769 (1992)
  [arXiv:hep-ph/9207218].
  
\bibitem{Barbier:2004ez}
  R.~Barbier, C.~Berat, M.~Besancon, M.~Chemtob, A.~Deandrea, E.~Dudas, P.~Fayet, S.~Lavignac {\it et al.},
  ``R-parity violating supersymmetry,''
  Phys.\ Rept.\  {\bf 420}, 1-202 (2005).
  [hep-ph/0406039].

\bibitem{Nath:2006ut}
  P.~Nath, P.~Fileviez P\'erez,
  ``Proton stability in grand unified theories, in strings and in branes,''
  Phys.\ Rept.\  {\bf 441}, 191-317 (2007).
  [hep-ph/0601023].

\bibitem{letter}
  P.~Fileviez~P\'erez, S.~Spinner and M.~K.~Trenkel,
  ``Testing the Mechanism for the LSP Stability at the LHC,''
  arXiv:1103.3824 [hep-ph].

\bibitem{PRL}
  P.~Fileviez P\'erez, S.~Spinner,
  ``Spontaneous R-Parity Breaking and Left-Right Symmetry,''
  Phys.\ Lett.\  {\bf B673 } (2009)  251-254.
  [arXiv:0811.3424 [hep-ph]];
  V.~Barger, P.~Fileviez P\'erez and S.~Spinner,
  ``Minimal gauged $U(1)_{B-L}$ model with spontaneous R-parity violation,''
  Phys.\ Rev.\ Lett.\  {\bf 102} (2009) 181802
  [arXiv:0812.3661 [hep-ph]].

\bibitem{Braun:2005nv}
  V.~Braun, Y.~-H.~He, B.~A.~Ovrut, T.~Pantev,
  ``The Exact MSSM spectrum from string theory,''
  JHEP {\bf 0605}, 043 (2006).
  [hep-th/0512177];
  M.~Ambroso, B.~Ovrut,
  ``The B-L/Electroweak Hierarchy in Heterotic String and M-Theory,''
  JHEP {\bf 0910}, 011 (2009).
  [arXiv:0904.4509 [hep-th]].
 
\bibitem{DeCampos:2010yu}
  F.~De Campos, O.~J.~P.~Eboli, M.~Hirsch, M.~B.~Magro, W.~Porod, D.~Restrepo and J.~W.~F.~Valle,
  ``Probing Neutrino Oscillations in Supersymmetric Models at the Large Hadron
  Collider,''
  Phys.\ Rev.\  D {\bf 82} (2010) 075002
  [arXiv:1006.5075 [hep-ph]].
  
\bibitem{Dreiner:2011xa}
  H.~K.~Dreiner, S.~Grab and T.~Stefaniak,
  ``Constraining Selectron LSP Scenarios with Tevatron Trilepton Searches,''
  arXiv:1103.1883 [hep-ph].
  
\bibitem{Bandyopadhyay:2010cu}
  P.~Bandyopadhyay, P.~Ghosh, S.~Roy,
  ``An unusual signal of Higgs boson in supersymmetry at the LHC,''
   [arXiv:1012.5762 [hep-ph]]. 
   
\bibitem{Mukhopadhyaya:2010qf}
  B.~Mukhopadhyaya, S.~Mukhopadhyay,
  ``Same-sign trileptons and four-leptons as signatures of new physics at the CERN Large Hadron Collider,''
  Phys.\ Rev.\  {\bf D82 } (2010)  031501.
  [arXiv:1005.3051 [hep-ph]].  
  
\bibitem{Mohapatra:1986aw}
  R.~N.~Mohapatra,
  ``Mechanism for understanding small neutrino mass in superstring theories,''
  Phys.\ Rev.\ Lett.\  {\bf 56}, 561 (1986).

\bibitem{Masiero:1990uj}
  A.~Masiero and J.~W.~F.~Valle,
  ``A model for spontaneous R-parity breaking,''
  Phys.\ Lett.\  B {\bf 251}, 273 (1990).

\bibitem{Takayama:2000uz}
  F.~Takayama, M.~Yamaguchi,
  ``Gravitino dark matter without R-parity,''
  Phys.\ Lett.\  {\bf B485}, 388-392 (2000).
  [hep-ph/0005214];
  W.~Buchmuller, L.~Covi, K.~Hamaguchi, A.~Ibarra, T.~Yanagida,
  ``Gravitino Dark Matter in R-Parity Breaking Vacua,''
  JHEP {\bf 0703}, 037 (2007).
  [hep-ph/0702184 [HEP-PH]].

\bibitem{Langacker}
  P.~Langacker,
  ``The Physics of Heavy Z-prime Gauge Bosons,''
  Rev.\ Mod.\ Phys.\  {\bf 81 } (2009)  1199-1228.
  [arXiv:0801.1345 [hep-ph]].
  
\bibitem{Basso:2010pe}
  L.~Basso, A.~Belyaev, S.~Moretti, G.~M.~Pruna and C.~H.~Shepherd-Themistocleous,
  ``$Z'$ discovery potential at the LHC in the minimal $B-L$ extension of the
  Standard Model,''
  arXiv:1002.3586 [hep-ph].

\bibitem{Huitu:2008gf}
  K.~Huitu, S.~Khalil, H.~Okada, S.~K.~Rai,
  ``Signatures for right-handed neutrinos at the Large Hadron Collider,''
  Phys.\ Rev.\ Lett.\  {\bf 101 } (2008)  181802.
  [arXiv:0803.2799 [hep-ph]].
  
\bibitem{AguilarSaavedra:2009ik}
  J.~A.~Aguilar-Saavedra,
  ``Heavy lepton pair production at LHC: Model discrimination with multi-lepton signals,''
  Nucl.\ Phys.\  {\bf B828 } (2010)  289-316.
  [arXiv:0905.2221 [hep-ph]]. 

\bibitem{TypeI-Zprime}
  P.~Fileviez P\'erez, T.~Han and T.~Li,
  ``Testability of Type I Seesaw at the CERN LHC: Revealing the Existence of
  the B-L Symmetry,''
  Phys.\ Rev.\  D {\bf 80} (2009) 073015
  [arXiv:0907.4186 [hep-ph]].

\bibitem{Fate}
  P.~Fileviez P\'erez, S.~Spinner,
  ``The Fate of R-Parity,''
  Phys.\ Rev.\  {\bf D83 } (2011)  035004.
  [arXiv:1005.4930 [hep-ph]].

\bibitem{Babu:2002tb}
  K.~S.~Babu, B.~Dutta, R.~N.~Mohapatra,
  ``Lepton flavor violation and the origin of the seesaw mechanism,''
  Phys.\ Rev.\  {\bf D67}, 076006 (2003).
  [hep-ph/0211068].

\bibitem{Rizzo:2006nw}
  T.~G.~Rizzo,
  ``$Z^\prime$ phenomenology and the LHC,''
  [hep-ph/0610104].

\bibitem{TypeI}
  P.~Minkowski,
  ``Mu $\to$ E Gamma At A Rate Of One Out Of 1-Billion Muon Decays?,''
  Phys.\ Lett.\ B {\bf 67} (1977) 421;
  T. Yanagida,
in {\it Proceedings of the Workshop on the Unified Theory
   and the Baryon Number in the Universe}, eds. O. Sawada et al.,
p.~95, KEK Report 79-18, Tsukuba (1979);
  M. Gell-Mann, P. Ramond and R. Slansky,
   in {\it Supergravity}, eds. P. van Nieuwenhuizen et al.,
   (North-Holland, 1979), p.~315;
  S.L. Glashow, in {\it Quarks and Leptons}, Carg\`ese, eds. M. L\'evy et al.,
(Plenum, 1980), p. 707;
  R.~N.~Mohapatra and G.~Senjanovi\'c,
  ``Neutrino Mass And Spontaneous Parity Nonconservation,''
  Phys.\ Rev.\ Lett.\  {\bf 44} (1980) 912.
  
\bibitem{Carena:2004xs}
  M.~S.~Carena, A.~Daleo, B.~A.~Dobrescu and T.~M.~P.~Tait,
  ``Z-prime gauge bosons at the Tevatron,''
  Phys.\ Rev.\  D {\bf 70} (2004) 093009
  [arXiv:hep-ph/0408098].
  

\bibitem{ggHiggsSUSY}
  A.~Djouadi,
  ``Squark effects on Higgs boson production and decay at the LHC,''
  Phys.\ Lett.\  B {\bf 435}, 101 (1998)
  [arXiv:hep-ph/9806315];
  M.~Spira,
  ``QCD effects in Higgs physics,''
  Fortsch.\ Phys.\  {\bf 46}, 203 (1998)
  [arXiv:hep-ph/9705337].

\bibitem{qqHiggsSUSY}
  A.~Dabelstein,
  ``Fermionic decays of neutral MSSM Higgs bosons at the one loop level,''
  Nucl.\ Phys.\  B {\bf 456}, 25 (1995)
  [arXiv:hep-ph/9503443];
  J.~A.~Coarasa Perez, R.~A.~Jimenez and J.~Sola,
  ``Strong effects on the hadronic widths of the neutral Higgs bosons in the
  MSSM,''
  Phys.\ Lett.\  B {\bf 389}, 312 (1996)
  [arXiv:hep-ph/9511402].

\bibitem{pdf}
  A.~D.~Martin, W.~J.~Stirling, R.~S.~Thorne and G.~Watt,
  ``Parton distributions for the LHC,''
  Eur.\ Phys.\ J.\  C {\bf 63} (2009) 189
  [arXiv:0901.0002 [hep-ph]].
  
\bibitem{delAguila:2008cj}
  F.~del Aguila, J.~A.~Aguilar-Saavedra,
  ``Distinguishing seesaw models at LHC with multi-lepton signals,''
  Nucl.\ Phys.\  {\bf B813}, 22-90 (2009).
  [arXiv:0808.2468 [hep-ph]]. 
  
\bibitem{Schwetz:2011qt}
  T.~Schwetz, M.~Tortola, J.~W.~F.~Valle,
  ``Global neutrino data and recent reactor fluxes: status of three-flavour oscillation parameters,''
  [arXiv:1103.0734 [hep-ph]].
  
\bibitem{Casas:2001sr}
  J.~A.~Casas, A.~Ibarra,
  ``Oscillating neutrinos and $\mu \to  e \gamma$,''
  Nucl.\ Phys.\  {\bf B618 } (2001)  171-204.
  [hep-ph/0103065].
  
  
\bibitem{Tao-DM}
  T.~Han, I.~-W.~Kim, J.~Song,
  ``Kinematic Cusps: Determining the Missing Particle Mass at Colliders,''
  Phys.\ Lett.\  {\bf B693 } (2010)  575-579.
  [arXiv:0906.5009 [hep-ph]].
  
\bibitem{DM-review}
  A.~J.~Barr, C.~G.~Lester,
  ``A Review of the Mass Measurement Techniques proposed for the Large Hadron Collider,''
  J.\ Phys.\ G {\bf G37 } (2010)  123001.
  [arXiv:1004.2732 [hep-ph]].

\bibitem{CMSboundGM}
  S.~Chatrchyan {\it et al.} [ CMS Collaboration ],
  ``Search for Supersymmetry in pp Collisions at sqrt(s) = 7 TeV in Events with Two Photons and Missing Transverse Energy,''
  [arXiv:1103.0953 [hep-ex]].
  
\bibitem{Khalil:2008ps}
  S.~Khalil, H.~Okada and T.~Toma,
  ``Right-handed Sneutrino Dark Matter in Supersymmetric B-L Model,''
  arXiv:1102.4249 [hep-ph].
  

\end{thebibliography}
\end{document}